\documentstyle[12pt]{article}

\ifx \doHeader\undefined \else
\doHeader{$Header: attable.tex,v 15.1 90/08/22 16:12:22 bxl Exp $}
\fi
\let\NEWtype=\relax
\def\NEW{\afterassignment\NEWexec\let\NEWtype=}
\def\NEWexec#1{\def\NEWobj{#1}\ifx#1\undefined\expandafter\doNEWtype\NEWobj\fi}
\def\SETNEW{\afterassignment\SETNEWexec\let\NEWtype=}
\def\SETNEWexec#1#2{\def\NEWobj{#1}%
  \ifx#1\undefined \expandafter\doNEWtype\NEWobj \if|#2|\else #1=#2\fi \fi}
\def\doNEWtype{\NEWtype}
\let\NEWtype=\relax

{\catcode`\P=1 \catcode`\T=2 \gdef\pt{PT}}
{\catcode`\E=1 \catcode`\X=2 \gdef\ex{EX}}
\global\futurelet\spacetoken{ }%
{\def\: {}\global\let\gulpspace=\:}
\let\ea=\expandafter
\def\debug#1{\begingroup\ifnum\tracingdebug>0 #1\fi\endgroup}
\def\wonline{\immediate\write16 }

\NEW\newcount\tracingdebug
\NEW\newif\ifresult
\NEW\newskip\filskip  \filskip =0pt plus 1fil
\NEW\newskip\fillskip \fillskip=0pt plus 1fill
\NEW\newskip\filhss   \filhss=0pt plus 1fil minus 1fil

\NEW\newcount\TMPcountA
\NEW\newcount\TMPcountB
\NEW\newcount\TMPcountC
\NEW\newcount\TMPcountD
\NEW\newcount\TMPcountE
\NEW\newdimen\TMPdimenA
\NEW\newdimen\TMPdimenB
\NEW\newdimen\TMPdimenC
\NEW\newdimen\TMPdimenD
\NEW\newdimen\TMPdimenE
\NEW\newskip\TMPskipA
\NEW\newbox\TMPboxA
\NEW\newbox\TMPboxB
\NEW\newtoks\TMPtoksA
\NEW\newtoks\TMPtoksB
\catcode`\@=11
\def\incr#1{\advance#1\@ne}
\def\decr#1{\advance#1\m@ne}
\ifx\undefined\Input \let\Input=\input \fi

\NEW\newcount\PubTeXuserstatus

\def\dimenassign@noop{\count255\z@}

\begingroup
\let\newmark=0 \let\newMark=0 \let\newcount=0 \let\newtoks=0
\gdef\setupnewmark{
  \let\ea=\expandafter       \let\nx=\noexpand
  \newcount\m@rkcount        \newtoks\m@rklist       \m@rklist={\relax}
  \outer\def\newmark##1{%
    \count@=\escapechar \escapechar=-1
    \edef\next{\global\m@rklist={\the\m@rklist\nx\nx\nx\or\join@##1\@}}\next
    \global\incr\m@rkcount 
    \edef\next{\the\m@rkcount}%
    \ea\ea\ea\xdef\join@\first##1{\nx\ifcase\next\nx\firstmark\nx\fi}%

\ea\ea\ea\xdef\join@\splitfirst##1{\nx\ifcase\next\nx\splitfirstmark\nx\fi}%
    \ea\ea\ea\xdef\join@\top##1{\nx\ifcase\next\nx\topmark\nx\fi}%
    \ea\ea\ea\xdef\join@\bot##1{\nx\ifcase\next\nx\botmark\nx\fi}%
    \ea\ea\ea\xdef\join@\splitbot##1{\nx\ifcase\next\nx\splitbotmark\nx\fi}%
    \xdef##1####1{\xdef\join@##1\@{####1}\nx\dummymark}%
    \ea\ea\ea\xdef\join@##1\@{}\escapechar=\count@}
  \def\join@##1##2{\csname\string##1\string##2\endcsname}
  \def\dummymark{{\let\\=0\ea\mark\ea{\the\m@rklist}}}%
  \newmark\tabheadmark
}
\gdef\setupnewMark{
	\newMark\tabheadmrk
	\def\tabheadmark{\Mark\tabheadmrk}
	\def\bottabheadmark{\Botmark\tabheadmrk}
	\def\toptabheadmark{\Topmark\tabheadmrk}
	\def\splitbottabheadmark{\Splitbotmark\tabheadmrk}
}
\endgroup

\ifx \Mark\undefined
	\let\next=\setupnewmark
\else
	\ifx \newMark\undefined
		\let\next=\setupnewmark
	\else
		\let\next=\setupnewMark
	\fi
\fi
\next

\ifx\undefined\alloc@@ 

\def\savecount{\alloc@@0\count\countdef\insc@unt}
\def\savedimen{\alloc@@1\dimen\dimendef\insc@unt}
\def\saveskip{\alloc@@2\skip\skipdef\insc@unt}
\def\savemuskip{\alloc@@3\muskip\muskipdef\@cclvi}
\def\savebox{\alloc@@4\box\chardef\insc@unt}
\def\savetoks{\alloc@@5\toks\toksdef\@cclvi}
\def\saveread{\alloc@@6\read\chardef\sixt@@n}
\def\savewrite{\alloc@@7\write\chardef\sixt@@n}
\def\savefam{\alloc@@8\fam\chardef\sixt@@n}

\def\alloc@@#1#2#3#4#5{\advance\count1#1by\@ne
  \ch@ck@#1#4#2
  \allocationnumber=\count1#1%
  #3#5=\allocationnumber
  \let\alloc@=\bad@alloc@ 
}
\def\ch@ck@#1#2#3{\ifnum\count1#1<#2%
  \else\bberror{No room to save a #3 register}\fi}

\let\orig@alloc@=\alloc@
\def\bad@alloc@{\relax\iftrue\message{\bb@badalloc}\fi\orig@alloc@}
\def\bb@badalloc{You should not allocate new registers while saving others!}

\fi 

\ifx \doHeader\undefined \else
\doHeader{$Header: atoutput.tex,v 15.1 90/08/22 16:12:02 bxl Exp $}
\fi

\catcode`\@=11

\toksdef\@ta=0 \toksdef\@tb=2
\gdef\outputstacktop{} 
\gdef\outputstackbot{} 

\newcount\tracingoutputroutines
\def\device{\ifnum\tracingonline>0 16 \else -1 \fi}
\def\wonline#1{\immediate\write\device{\outputdepth #1}} 

\def\newoutput{%
  \@ta=\expandafter{\outputstacktop}%
  \xdef\outputstacktop{\the\output{\the\@ta}}%
  \afterassignment\modifyoutput
  \@ta
}

\def\modifyoutput{%
  \global\output=\expandafter{\expandafter\outputhead\the\@ta\outputtail}%
}

\def\outputhead{\debugoutput \upoutput}
\edef\singledot{.}
\let\outputfinish=\empty
\def\outputtail{\downoutput\ifx\outputdepth\singledot\outputfinish\fi}

\def\endnewoutput{%
  \afterassignment\newoutputstacktop
  \global\output=\outputstacktop{}{}%
}

\def\upoutput{
  \@ta=\expandafter{\outputstackbot}%
  \xdef\outputstackbot{\the\output{\the\@ta}}%
  \afterassignment\newoutputstacktop
  \global\output=\outputstacktop{}{}%
}

\def\downoutput{
  \@ta=\expandafter{\outputstacktop}%
  \xdef\outputstacktop{\the\output{\the\@ta}}%
  \afterassignment\newoutputstackbot
  \global\output=\outputstackbot{}{}%
}

\def\newoutputstacktop{\gdef\outputstacktop}
\def\newoutputstackbot{\gdef\outputstackbot}

\def\shazam{
  \afterassignment\@tb
  \@ta=\outputstackbot{}{}%
  \afterassignment\modifyoutputstackbot
  \@ta
}
\def\modifyoutputstackbot{%
  \xdef\outputstackbot{%
    {\noexpand\outputhead\the\@ta\noexpand\outputtail}{\the\@tb}%
  }
}

\def\sepuku{\let\downoutput=\ukupes}

\def\ukupes{
  \afterassignment\newoutputstackbot
  \@ta=\outputstackbot{}{}%
}

\def\debugoutput{%
  \setuptracingoutputroutines
  \ifnum\tracingoutputroutines>0
   \bgroup
    \edef\device{\device}
    \wonline{TRACING OUTPUT ->\the\output.}
    \wonline{ Output penalty->\the\outputpenalty.}
    \ifnum\tracingoutputroutines>1
      \wonline{Tracing marks:}
      \toks0=\expandafter{\topmark}   \wonline{ topmark -->\the\toks0 .}
      \toks0=\expandafter{\firstmark} \wonline{ firstmark->\the\toks0 .}
      \toks0=\expandafter{\botmark}   \wonline{ botmark -->\the\toks0 .}
      \wonline{ ht of box 255 ->\the\ht255 .}
      \ifnum\tracingoutputroutines>2
        \wonline{Box 255:}
        \scrollmode\showbox255 \errorstopmode
      \fi
    \fi
   \egroup 
   \tracingoutputroutines=1 
  \fi
}

\def\setuptracingoutputroutines{
  \edef\outputdepth{\outputdepth .}%
}
\def\outputdepth{}

\def\TabRuleVI{-} \def\TabRuleVO{-}
\def\TabRuleHI{-} \def\TabRuleHO{-}
\def\TabRuleHH{-}
\def\TabJustVO{t} \def\TabJustVH{t} \def\TabJustHO{c}
\def\TabW{} \def\TabOffset{} \def\TabCellHt{}
\ifx\undefined\TableWd		\def\TableWd{Z}		\fi
\ifx\undefined\TableWdAAmt	\def\TableWdAAmt{\hsize}\fi
\ifx\undefined\TableJust	\def\TableJust{X}	\fi
\ifx\undefined\TableJustLAmt	\def\TableJustLAmt{0pt }\fi

\SETNEW\newcount\TableWdRPct{100}
\NEW\newdimen\curtypesize

\ifx \settolerance\undefined \let\settolerance=\relax \fi

\NEW\newif\ifinTableCell		\inTableCellfalse

\NEW\newdimen\@sentinel	\@sentinel=-1000pt 

\NEW\newdimen\Fig@width	\Fig@width=\@sentinel
\NEW\newdimen\Fig@window	\Fig@window=\@sentinel
\NEW\newskip\Fig@leftskip	\Fig@leftskip=\@sentinel
\NEW\newskip\Fig@rightskip	\Fig@rightskip=\@sentinel

\def\@blank{.}
\def\@single{-}
\def\@double{=}
\def\@bold{+}
\def\@top{t}
\def\@bot{b}
\def\@center{c}
\def\@both{b}
\def\@left{l}
\def\@right{r}
\def\@default{*}

\NEW\newdimen\Tab@width
\NEW\newdimen\Tab@window
\NEW\newskip\Tab@leftskip
\NEW\newskip\Tab@rightskip

\NEW\newcount\TabColC	\TabColC=1
\NEW\newdimen\Tab@wd
\NEW\newdimen\Tab@kern

\NEW\newcount\column
\NEW\newcount\row

\NEW\newbox\rrule@box
\NEW\newbox\row@box
\NEW\newbox\cell@box
\NEW\newbox\xrow@box
\NEW\newbox\xrow@boxsofar
\NEW\newbox\brk@rule
\NEW\newbox\row@migrations
\NEW\newbox\row@RVBmigrations

\SETNEW\newcount\interrowpenalty{50}
\SETNEW\newcount\splitrowpenalty{10000}
\SETNEW\newcount\clubrowpenalty{5000}
\SETNEW\newcount\almostclubrowpenalty{2000}
\SETNEW\newcount\widowrowpenalty{10000}
\SETNEW\newcount\postheadingpenalty{10000}

\NEW\newcount\tabhead@number
\NEW\newbox\tabheadboxA
\NEW\newbox\tabheadboxB

\NEW\newif\ifduplicatinglastrule \duplicatinglastruletrue
\NEW\newif\ifdonewwindow \global\donewwindowfalse

\def\Table@ifpreheading{\ifx\next\Tabheading \resulttrue\fi
	\ifnum\row=\z@\ifintableheading \resulttrue\fi\fi}
\def\Table@ifpostheading{\ifx\next\Endtabheading \resulttrue\fi
        \ifnum\row=0 \ifusingheading \resulttrue \fi\fi}
\def\Table@iftabletop{\ifnum\row=0 \ifusingheading\else \resulttrue \fi\fi}
\def\Table@iftablebot{\ifx\next\Endtable \resulttrue\fi}
\def\Table@iftableleft{\ifnum\column=0 \resulttrue\fi}
\def\Table@iftableright{\ifnum\column=\TabColC \resulttrue\fi}

\def\savebaselineskip{
  \ifdim\lineskiplimit<\maxdimen
    \normalbaselineskip =\baselineskip
    \normallineskiplimit=\lineskiplimit
    \normallineskip     =\lineskip
  \fi
}

\def\Tbl[#1]{\begingroup\ignorespaces #1}
\def\EndTbl[]{\endgroup}

\let\tablestart=\relax

\def\Table#1[#2]{%
  \endgraf 				
  \global\let\TblInsert=T%
  \begingroup
    \ifinTableCell
      \nointerlineskip	
    \else
      \ifdim\prevdepth>-1000pt
        \TMPskipA=\baselineskip
	\advance\TMPskipA by-\prevdepth
	\ifdim\TMPskipA>\lineskiplimit
	  \vskip\TMPskipA
	\else
	  \vskip\lineskip
	\fi
      \fi
      {\setbox0=\vbox to12pt{}\nointerlineskip\box0\kern-12pt}%
      \ifnum\TableRuleWt>\z@ \table@setrules \fi
      \let\adjustbaselineskip=\empty 
      \tablestart
    \fi
    \tabheadmark{0}%
    \global\setbox\ifodd\tabhead@number\tabheadboxA\else\tabheadboxB\fi=%
      \box\voidb@x
    \Table@dflt 
    \ignorespaces#2%
    \Table@fonts
    \Table@
}

\def\Endtable#1[#2]{%
    \ifnum\Rpenalty<\maxdimen \else
      \Rpenalty=\widowrowpenalty
    \fi
    \put@row\Rpenalty
    \ignorespaces #2\relax
  \endgroup
  \ifinTableCell \nointerlineskip	  	
  \else \par\fi
  \tabheadmark{}%
}

\def\table@setrules{%
      \ifcase\TableRuleWt
      \or
        \SingleRuleWidthInPixels=2
      \or
        \SingleRuleWidthInPixels=4
      \or
        \SingleRuleWidthInPixels=6
      \or
        \SingleRuleWidthInPixels=8
      \or
        \SingleRuleWidthInPixels=4
      \or
        \SingleRuleWidthInPixels=4
      \or
        \SingleRuleWidthInPixels=4
      \or
        \SingleRuleWidthInPixels=4
      \else
        \rule@th=\TableRuleWtAmt
      \fi
      \ifnum\TableRuleWt<9
        \rule@th=\SingleRuleWidthInPixels in
        \divide\rule@th by 300
      \fi
}

\def\Table@fonts{%
    \TableOther@fnt \ea\let\ea\CsA\the\font \let\Tab@ord@font=\CsA
    \nullfont \tracinglostchars=1 
    \spaceskip=0pt 
    \savebaselineskip\offinterlineskip
}

\def\TableOther@fnt{\relax
  \ifx \alphatypesize\undefined \else
    \let\otherfontfam=\TableFntFfmOtherName
    \TableFntFfm
    \settypesize{\TableFntTsz}{\TableFntTszPercentmain}%
      {\TableFntTszIncrmain}{\TableFntTszOtherSize}%
    \TableFntTst
  \fi
}

\def\Table@dflt{
    \ifinTableCell
      \Tab@leftskip=\z@
      \Tab@rightskip=\z@
      \Tab@width=\hsize
    \else
      \Tab@leftskip=\leftskip
      \Tab@rightskip=\rightskip
      \Tab@width=\hsize
      \advance\Tab@width by-\Tab@leftskip
      \advance\Tab@width by-\Tab@rightskip
    \fi
    \def\TabRuleVS{}
    \def\TabJustHS{}
    \TabColC=\z@			
    \def\TabW{}
    \def\TabOffset{}
    \def\TabColW{}
    \def\TabCellHt{}
    \cell@ht=\@sentinel			
    \row=-1 
    \global\setbox\xrow@box=\box\voidb@x
    \intableheadingfalse
    \usingheadingfalse
    \Table@dflt@other 
}

\ifx\undefined\Table@dflt@other
    \def\Table@dflt@other{
    }
\fi

\def\Table@{%
    \ifx\empty\TabCellHt \else
      \TMPdimenA=\fontdimen5\font
      \fontdimen5\font=\trial@unit \cell@ht=\TabCellHt\ex
      \fontdimen5\font=\TMPdimenA
      \let\TabCellHt=\empty
    \fi
    \ifinTableCell\else			
      \ifinner
   	\Tab@window=\@sentinel
      \else				
   	\ifdim\pagegoal<\maxdimen	
          \Tab@window=\pagegoal \advance\Tab@window by-\pagetotal
   	\else				
          \Tab@window=\vsize
   	\fi
      \fi
    \fi
    \ifnum\interrowpenalty  > 9999	
      \tab@nosplitting
    \else				
      \ifinTableCell \tab@allowsplitting \fi
    \fi
    \dimen@=\fontdimen5\font    
    \ifx\TabW\empty \else       
      \TMPdimenA=\Tab@width
      \Tab@width=\TabW\TMPdimenA\dimenassign@noop       
    \fi
    \ifx\TabOffset\empty \else  
      \TMPdimenA=\Tab@width				 
      \advance\Tab@leftskip by\TabOffset\TMPdimenA\dimenassign@noop
      \Tab@rightskip=\filhss
    \fi
    \ifinTableCell\else
      \if A\TableWd\relax
        \Tab@width=\TableWdAAmt 
      \else
        \if R\TableWd\relax
          \ifnum\TableWdRPct=100 \else
            \Tab@width=.01\Tab@width
            \multiply\Tab@width by\TableWdRPct 
          \fi
        \fi
      \fi
      \ifdim\Tab@width<10pt
        \message{Table: overall table width requested is \the\Tab@width;
          using 10pt.}%
        \Tab@width=10pt
      \fi
    \fi
    \ifx\TabColW\empty		
	\Table@I \fi		
    \Table@II			
    \ifdim\Tab@wd=\TMPdimenA	
        \Tab@width=\Tab@wd	
    \else			
	\Table@III		
        \ifdim\Tab@kern>.2\p@	
	  \Table@IV		
        \fi			
    \fi				
    \Table@V			
    \Table@VI 
    \ifinTableCell\else
      \TMPdimenA=\hsize
      \advance\TMPdimenA-\Tab@width
      \advance\TMPdimenA-\Tab@leftskip
      \advance\TMPdimenA-\Tab@rightskip
      \if L\TableJust\relax
         \advance\Tab@leftskip  \TableJustLAmt
         \advance\Tab@rightskip \TMPdimenA
         \advance\Tab@rightskip-\TableJustLAmt
      \else\if R\TableJust\relax
         \advance\Tab@leftskip  \TMPdimenA
      \else\if C\TableJust\relax
         \advance\Tab@leftskip  .5\TMPdimenA
         \advance\Tab@rightskip .5\TMPdimenA
      \else
         \advance\Tab@rightskip \TMPdimenA
      \fi\fi\fi
    \fi
    \initdecimalalign           
    \fontdimen5\font=\dimen@	
    \crule@setup
    \tab@init@allcols
}

\def\tab@nosplitting{%
	\splitrowpenalty   = 10000
	\clubrowpenalty    = 10000
	\widowrowpenalty   = 10000
	\postheadingpenalty= 10000
}

\def\tab@allowsplitting{%
  \ifnum\splitrowpenalty   <-9999	\splitrowpenalty    =-9999 \fi
  \ifnum\clubrowpenalty    > 9999	\clubrowpenalty     = 9999 \fi
  \ifnum\widowrowpenalty   > 9999	\widowrowpenalty    = 9999 \fi
  \ifnum\postheadingpenalty> 9999	\postheadingpenalty = 9999 \fi
}

\def\tab@init@allcols{%
  \column=0
  \loop
    \incr\column
    \tab@init@column\column\column
  \ifnum\column<\TabColC\repeat
}

\def\tab@init@column#1#2{%
  \begingroup
    \cell@kern=\z@ \let\\=\origOr
    \cell@wd=\z@
    \TMPcountB=#1\relax
    \loop
      \advance\cell@wd\ifcase\number#1\TabColW@\fi\pt
    \ifnum\TMPcountB<#2\relax
      \incr\TMPcountB
    \repeat
    \edef\TabJustHL{\ifcase\number#1\TabJustHS\else*\fi}%
    \ifx\TabJustHL\@default
	\let\TabJustHL=\TabJustHO
	\leftskip=\TabGutL 
	\ifx\TabJustHL\@both\else
	\ifx\TabJustHL\@left\else
	    \advance\leftskip \fillskip
	\fi\fi
    \fi
    \edef\TabJustHR{\ifcase\number#2\TabJustHS\else*\fi}%
    \ifx\TabJustHR\@default
	\let\TabJustHR=\TabJustHO
	\rightskip=\TabGutR 
	\ifx\TabJustHR\@both\else
	\ifx\TabJustHR\@right\else
	    \advance\rightskip \fillskip
	\fi\fi
    \fi
    \Tab@ord@font 
    \hsize=\cell@wd 
    \loosenup 
    \xdef\GCsA{%
      \cell@kern=\z@
      \cell@wd=\the\cell@wd
      \edef\nx\TabJustHL{\TabJustHL}%
      \edef\nx\TabJustHR{\TabJustHR}%
      \leftskip=\the\leftskip
      \rightskip=\the\rightskip
      \spaceskip=\the\spaceskip
      \xspaceskip=\the\xspaceskip
      \relax
    }%
  \endgroup
  \ea\let\csname cell@init@\number#1-\number#2\endcsname=\GCsA
  \ea\savedimen\csname VSpankern\number#1\endcsname 
  \csname VSpankern\number#1\endcsname=\z@
}

\def\Table@I{%
    \ifnum\TabColC<1
	\message{Table error: Neither \noexpand\TabColW\space nor
	    \TabColC\space have been given legal values.
	    I'm assuming one full width column.}%
	\TabColC=1
	\def\TabColW{\\1}%
    \else
	\TMPdimenA=\Tab@Width \divide\TMPdimenA by\TabColC
	\let\\=0\TMPcountA=\TabColC
	\loop\ifnum\TMPcountA>0
	    \edef\TabColW{\TabColW\\\the\TMPdimenA}%
	    \decr\TMPcountA
	\repeat
    \fi
}

\def\Table@II{%
    \fontdimen5\font=\z@			
    \def\\{\ex\incr\TMPcountA\advance\TMPdimenA}%
    \TMPcountA=\z@ \TMPdimenA=\z@ \TabColW\ex	
    \fontdimen5\font=\trial@unit		
    \def\\{\ex\advance\Tab@wd}%
    \Tab@wd=\z@ \TabColW\ex			
    \Tab@kern=\Tab@width \advance\Tab@kern-\Tab@wd
}

\NEW\newdimen\trial@unit \trial@unit=.25pt	
\NEW\newdimen\halfmaxdimen
\halfmaxdimen=\maxdimen \divide\halfmaxdimen by2

\def\Table@III{%
    \TMPdimenB=\Tab@wd
    \advance\TMPdimenB-\TMPdimenA	
    \TMPdimenC=\Tab@width \advance\TMPdimenC-\TMPdimenA	
    \TMPcountC=1 
    \loop\ifdim\TMPdimenB>2\trial@unit
	\divide\TMPdimenB by 2 
	\advance\TMPcountC \TMPcountC
    \repeat
    \TMPcountE=1 
    \loop\ifdim\TMPdimenB<\trial@unit
	\multiply\TMPdimenB by 2 
	\advance\TMPcountE \TMPcountE
    \repeat
    \loop\ifdim\TMPdimenC<\halfmaxdimen
	\multiply\TMPdimenC by 2 
	\advance\TMPcountC \TMPcountC
    \repeat
    \divide\TMPdimenC by \TMPdimenB	
    \multiply\TMPdimenC \trial@unit	
    \ifnum\TMPcountC>\TMPcountE
	\divide\TMPcountC by \TMPcountE
	\divide\TMPdimenC by \TMPcountC	
    \else
	\divide\TMPcountE by \TMPcountC
	\multiply\TMPdimenC by \TMPcountE
    \fi
    \def\\{\ex\advance\Tab@wd}%
    \fontdimen5\font=\TMPdimenC 
    \Tab@wd=\z@ \TabColW\ex		
    \Tab@kern=\Tab@width \advance\Tab@kern-\Tab@wd	
    \trial@unit=\TMPdimenC
}

\def\Table@IV{%
    \def\\{\ex\advance\Tab@wd}%
    \ifdim\Tab@wd<\Tab@width \TMPdimenE=1sp \let\CsA=\testup
		       \else \TMPdimenE=-1sp\let\CsA=\testdown\fi
    \resulttrue
    \loop
	\advance\TMPdimenC\TMPdimenE	
	\fontdimen5\font=\TMPdimenC	
	\Tab@wd=\z@ \TabColW\ex		
	\advance\TMPdimenE\TMPdimenE	
	\CsA
    \ifresult\repeat
    \TMPdimenE=-.5\TMPdimenE		
    \resulttrue
    \loop
        \ifdim\TMPdimenE=\z@ \resultfalse\fi	
    \ifresult
	\advance\TMPdimenC\TMPdimenE		
	\fontdimen5\font=\TMPdimenC 		
	\Tab@wd=\z@ \TabColW\ex 		
	\TMPdimenE=
	  \ifdim\Tab@wd>\Tab@width
	    \ifdim\TMPdimenE>\z@ -\fi 		
	  \else
	    \ifdim\TMPdimenE<\z@ -\fi 		
	  \fi
	  .5\TMPdimenE 				
    \repeat
    \Tab@kern=\Tab@width \advance\Tab@kern-\Tab@wd
    \trial@unit=\TMPdimenC
}

\def\testup{\ifdim\Tab@wd>\Tab@width \resultfalse\fi}
\def\testdown{\ifdim\Tab@wd<\Tab@width \resultfalse\fi}

\NEW\newif\ifrowtoowide

\def\Table@V{
    \fontdimen5\font=\trial@unit
    \let\\=\Table@V@I 
    \Tab@wd=\z@ \def\TabColW@{}\TabColW\ex
    \Tab@kern=\Tab@width \advance\Tab@kern-\Tab@wd
    \ifdim\Tab@kern<\z@ \rowtoowidetrue \else \rowtoowidefalse\fi 
}

\def\Table@V@I{\ex\afterassignment\Table@V@II\TMPdimenA=}%
\def\Table@V@II{%
    \advance\Tab@wd\TMPdimenA
    \toks0=\ea{\TabColW@\\}\edef\TabColW@{\the\toks0 \the\TMPdimenA}%
}

\def\Table@V@III{\ex\afterassignment\Table@V@IV\TMPdimenA=}%
\def\Table@V@IV{
    \advance\TMPdimenA\Tab@kern \Tab@kern=\z@ \let\\=\Table@V@I
    \advance\Tab@wd\TMPdimenA
    \toks0=\ea{\TabColW@\\}\edef\TabColW@{\the\toks0 \the\TMPdimenA}%
}

\def\Table@VI{%
    \ifnum\TMPcountA=\TabColC \else
	\ifnum\TabColC=\z@
	    \TabColC=\TMPcountA
	\else
	    \message{Table: Warning; Table column count (\the\TabColC) does
		not agree with the number of columns in the width specification
		(\the\TMPcountA).  Using the latter.}%
	\fi
    \fi
}

\NEW\newdimen\rule@th
\def\ruleth{\rule@th}

\NEW\newcount\ResolutionDPI
\NEW\newcount\SingleRuleWidthInPixels
\ifx \TableRuleWt\undefined \def\TableRuleWt{3}\fi
\NEW\newdimen\TableRuleWtAmt

\NEW\newtoks\aftertabheadtoks 
\def\aftertabhead #1{\aftertabheadtoks=\ea{\the\aftertabheadtoks #1}}

\rule@th=\SingleRuleWidthInPixels in
\divide\rule@th by\ResolutionDPI

\def\Rrule#1[#2]{\Rrule@setup\ignorespaces#2\Rrule@I}

\def\Rrule@setup{%
    \let\TabRuleH=\@default
    \let\TabRuleHS=\empty
    \let\TabRuleSave=\empty
    \let\TabRuleUse=\empty
}

\def\Rrule@I {%
    \futurelet\next\Rrule@Ia}
\def\Rrule@Ia{%
    \ifx\next\spacetoken
        \global\let\CsG=\Rrule@Ib
    \else
        \global\let\CsG=\Rrule@Ic
    \fi\CsG}
\def\Rrule@Ib{%
    \ea\Rrule@I\gulpspace}
\def\Rrule@Ic{%
  \ifx\TabRuleUse\empty
    \ifx\TabRuleH\@default		
	\resultfalse
	\Table@ifpreheading \ifresult \let\TabRuleH=\TabRuleHH \else
	\Table@ifpostheading\ifresult \let\TabRuleH=\TabRuleHH \else
	\Table@iftabletop   \ifresult \let\TabRuleH=\TabRuleHO \else
	\Table@iftablebot   \ifresult \let\TabRuleH=\TabRuleHO \else 
				      \let\TabRuleH=\TabRuleHI
	\fi\fi\fi\fi
    \fi
    \ifx\TabRuleHS\empty
	\let\next=\Rrule@III		
    \else
	\let\next=\Rrule@IV		
    \fi
  \else
    \let\next=\Rrule@VI			
  \fi
  \next
  \ifx\TabRuleSave\empty \else
    \edef\next{\csname @bx\TabRuleSave\endcsname}%
    \ea\savebox\next
    \global\setbox\next=\copy\rrule@box
    \ifintableheading
      \edef\next{{\nx\savebox\next\global\setbox\next=\box\number\next\relax}}%
      \ea\aftertabhead\next
    \fi
  \fi
}

\def\Rrule@III{%
    \setbox\rrule@box=\hbox 
    {
	\kern\Tab@leftskip
	\Rrule@draw\TabRuleH \Tab@wd
	\ifrowtoowide \kern\Tab@kern \fi
    }%
}

\def\Rrule@IV{%
    \setbox\rrule@box=\hbox 
    {
	\kern\Tab@leftskip
	\let\\=\Rrule@V
	\column=\z@ \TabColW@
	\ifrowtoowide \kern\Tab@kern \fi
    }%
}

\def\Rrule@V{%
    \ex\let\\=\or
    \incr\column
    \edef\next{\ifcase\the\column\TabRuleHS\fi}
    \let\\=\Rrule@V
    \ifx\next\@default
	\Rrule@draw\TabRuleH
    \else
	\Rrule@draw\next
    \fi
}

\def\Rrule@VI{%
    \setbox\rrule@box=\copy\csname @bx\TabRuleUse\endcsname
}

\def\Rrule@draw #1{
    \ifx#1\@blank \rrule@bl \else
    \ifx#1\@single\rrule@si \else
    \ifx#1\@double\rrule@do \else
    \ifx#1\@bold  \rrule@bo \else
    \rrule@bl \fi\fi\fi\fi
}

\def\rrule@ #1{%
    \hbox to\TMPdimenA{%
	\kern-.5\rule@th
	\leaders\hrule height#1\hfill
	\kern-.5\rule@th}}

\def\rrule@bl{\afterassignment\rrule@bl@ \TMPdimenA=}
\def\rrule@bl@{%
    \vbox to \z@{\hbox to\TMPdimenA{\hss}}}

\def\rrule@si{\afterassignment\rrule@si@ \TMPdimenA=}
\def\rrule@si@{%
    \vbox to \z@{%
	\vss
	\rrule@{\rule@th}%
	\vss}}

\def\rrule@do{\afterassignment\rrule@do@ \TMPdimenA=}
\def\rrule@do@{%
    \vbox to \z@{%
	\vss
	\rrule@{.5\rule@th}%
	\kern\rule@th\nointerlineskip
	\rrule@{.5\rule@th}%
	\vss}}

\def\rrule@bo{\afterassignment\rrule@bo@ \TMPdimenA=}
\def\rrule@bo@{%
    \vbox to \z@{%
	\vss
	\rrule@{2\rule@th}%
	\vss}}

\NEW\newif\ifvcenteringrow
\NEW\newif\iffixedhtrow
\NEW\newif\ifTabNobreak
\NEW\newcount\Rpenalty \Rpenalty=\maxdimen 

\ifx\undefined\RowLeftExec \let\RowLeftExec=\empty \fi

\def\Row#1[#2]{%
    \ifnum\Rpenalty<\maxdimen
    \else
      \Rpenalty =
	\ifcase\row \postheadingpenalty
	\or \clubrowpenalty
	\or \almostclubrowpenalty
	\else
	    \ifnum\row<0 10000 
	    \else \interrowpenalty \fi
	\fi
    \fi
    \put@row\Rpenalty
    \Rpenalty=\maxdimen
    \ifTabNobreak \ifnum\skip@row@penalties<1 \skip@row@penalties=1 \fi\fi
    \setbox\row@box=\hbox 
    \bgroup
        \needmigrationtrue
	\adjustinto\row@migrations
	\RVBadjustinto\row@RVBmigrations
	\RowLeftExec 
	\let\TabJustV=\@default
	\ignorespaces #2%
	\ifx \TabJustV\@default
		\ifintableheading
			\let\TabJustV=\TabJustVH
		\else
			\let\TabJustV=\TabJustVO
		\fi
	\fi
	\column=\z@
	\row@defaults@for@cells 
	\ifdim\cell@ht=\@sentinel\else
	  \setbox\TMPboxA=\hbox{}\ht\TMPboxA=\cell@ht\box\TMPboxA
	  \fixedhtrowtrue 
	\fi
	\kern\Tab@leftskip
        \ignorespaces
}

\def\row@defaults@for@cells{%
  \row@setcellht
  \row@setcellflags
  \row@setfont
  \row@settopkern
  \def\VSpanData{}%
}

\def\row@setvtopbox #1{%
    \let\vtopbox=\vbox 
}

\def\row@setcellht{%
    \ifx\empty\TabCellHt
    \else
      \cell@ht=\TabCellHt\relax 
      \let\TabCellHt=\empty
    \fi
}

\def\row@setfont{%
    \ifintableheading
      \Tab@chd@font \normalbaselines
    \else
      \Tab@ord@font \normalbaselines 
    \fi
}

\def\row@setcellflags{%
    \ifdim\cell@ht=\@sentinel \fixedhtrowfalse\else\fixedhtrowtrue\fi
}

\def\row@settopkern{%
      \cell@topkern=\TabGutT
}

\def\Endrow#1[#2]{%
        \ignorespaces #2%
	\ifrowtoowide \kern\Tab@kern \fi		
	\global\let\GCsA=\VSpanData
	\global\TMPcountD=\Rpenalty
    \egroup				
    \Rpenalty=\TMPcountD
    \table@fixcellheights\row@box 
    \TabNobreakfalse 
    \GCsA 
    \ifdim\Tab@window=\@sentinel
    \else
	\advance\Tab@window-\ht\row@box
	\advance\Tab@window-\dp\row@box
	\advance\Tab@window-\ht\xrow@box
	\advance\Tab@window-\dp\xrow@box
	\advance\Tab@window \rule@th 
        \ifdonewwindow \global\donewwindowfalse
	  \resulttrue
	  \ifinner \resultfalse \fi
	  \ifdim\pagegoal=\maxdimen \resultfalse \fi
          \ifresult
	    \Tab@window=\pagegoal
	    \advance\Tab@window-\pagetotal
	  \else
	    \Tab@window=\vsize
	  \fi
        \fi
    \fi
    \incr\row
    \ignorespaces
}

\NEW\newcount\skip@row@penalties 
\ifx\undefined\putrowExec \let\putrowExec=\empty \fi

\NEW\newbox\lastrrule@box 

\def\put@row #1{%
    \endgraf
    \ifnum\skip@row@penalties>0
        \decr\skip@row@penalties
    \else
        \penalty #1
    \fi
    \putrowExec
    \ifdim\ht\row@box>-\dp\row@box	
	\nointerlineskip
	\box\row@box
    \else
	\setbox\row@box=\box\voidb@x	
    \fi
    \ifvbox\row@migrations
        \outeradjust{\unvbox\row@migrations\nointerlineskip}%
    \fi
    \ifvbox\row@RVBmigrations
        \RVBdropmigrations{\box\row@RVBmigrations}%
    \fi
    \nointerlineskip
    \setbox\lastrrule@box=\copy\rrule@box
    \box\rrule@box			
    \endgraf
    \ifdim\Tab@window=\@sentinel \else
	\TMPdimenA=\pagegoal \advance\TMPdimenA-\pagetotal
	\advance\TMPdimenA\pageshrink
	\ifdim\TMPdimenA<\z@ \penalty9999 \Tab@window=\z@ \fi
    \fi
}

\def\VSpanUpdate #1#2{%
  \ifnum#2>\@ne
    \TMPdimenA=\ht\row@box
    \advance\TMPdimenA by\dp\row@box
    \ea\advance\csname VSpankern\number#1\endcsname by\TMPdimenA
    \TabNobreaktrue 
  \else
    \csname VSpankern\number#1\endcsname=\z@
  \fi
}

\def\crskip #1{\hskip#1\rule@th plus1sp minus 1sp}
\def\crrule #1{\vrule width#1\rule@th}

\NEW\newbox\crule@bl@box
\NEW\newbox\crule@si@box
\NEW\newbox\crule@do@box
\NEW\newbox\crule@bo@box

\def\crule@setup{%
  \setbox\crule@bl@box=\hbox{\crule@bl}%
  \setbox\crule@si@box=\hbox{\crule@si}%
  \setbox\crule@do@box=\hbox{\crule@do}%
  \setbox\crule@bo@box=\hbox{\crule@bo}%
}

\edef\crule@bl{\crskip{0}}
\edef\crule@si{\crskip{-.5}\crrule{1}\crskip{-.5}}
\edef\crule@do{\crskip{-1}\crrule{.5}\crskip{1}\crrule{.5}\crskip{-1}}
\edef\crule@bo{\crskip{-1}\crrule{2}\crskip{-1}}

\def\Crule@draw#1{
  \copy
    \ifx#1\@blank \crule@bl@box \else
    \ifx#1\@single\crule@si@box \else
    \ifx#1\@double\crule@do@box \else
    \ifx#1\@bold  \crule@bo@box \else
		  \crule@bl@box 
    \fi\fi\fi\fi
}

\def\Crule#1[#2]{%
    \let\\=\or \edef\TabRuleV{\ea\ifcase\ea\column\TabRuleVS\fi}%
    \ifx\TabRuleV\empty \let\TabRuleV=\@default\fi
    \ifx\TabRuleV\@default \edef\TabRuleV{\TabRuleIOtest}\fi
    \ignorespaces #2%
    \Crule@draw\TabRuleV
    \ignorespaces
}

\def\TabRuleIOtest{%
    \ifnum\column=\z@		\TabRuleVO \else
    \ifnum\column=\TabColC	\TabRuleVO \else
				\TabRuleVI
    \fi\fi
}

\NEW\newcount\last@column
\NEW\newdimen\cell@wd
\NEW\newdimen\cell@ht
\NEW\newdimen\cell@kern
\NEW\newdimen\cell@topkern

\NEW\newcount\TabColSpan
\NEW\newcount\TabVSpan

\SETNEW\newskip\TabGutT{5pt minus 5pt}
\SETNEW\newskip\TabGutB{5pt minus 5pt}
\SETNEW\newskip\TabGutL{5pt}
\SETNEW\newskip\TabGutR{5pt}

\NEW\newif\ifsplitallowed
\NEW\newif\ifallowoverfullcells \allowoverfullcellstrue

\def\caphtoverex{1.55}

\ifx\undefined\CellTopExec \let\CellTopExec=\empty \fi

\def\Cell#1[#2]{%
    \incr\column
    \TabColSpan=1
    \TabVSpan=1
    \def\TabJustH{*}%
    \def\CellShade{0}%
    \ignorespaces #2%
    \last@column=\column 
    \ifnum\TabColSpan>1 \cell@init@spanned \else \cell@init@unspanned \fi
    \TMPcountA=\column
    \loop
      \toks0=\ea{\VSpanData}%
      \edef\VSpanData{%
      	\the\toks0 \nx\VSpanUpdate{\the\TMPcountA}{\the\TabVSpan}}%
    \ifnum\TMPcountA<\last@column
      \incr\TMPcountA
    \repeat
    \setbox\cell@box=\vbox
    \iffixedhtrow\ifallowoverfullcells to\cell@ht \fi\fi
    \bgroup
      \ifnum\TabVSpan>1
      \else
	\Cell@II		
	\Fig@leftskip=\z@	
	\Fig@rightskip=\z@
	\Fig@width=\cell@wd
	\TMPdimenA=\csname VSpankern\number\column\endcsname
	\ifdim\TMPdimenA>\z@
	  \kern-\TMPdimenA
	\fi
	\TMPskipA=\TabGutT
	\ifx\TabJustV\@top
	\else
	  \advance\TMPskipA by 0pt plus 1fil
	  \relax
	\fi
        \vskip\TMPskipA 
	\TMPdimenB=\baselineskip
	\advance\TMPdimenB by-\caphtoverex\fontdimen5\font
	\prevdepth=\TMPdimenB
	\inTableCelltrue
	\settolerance
      \fi
      \ignorespaces
}

\def\cell@init@spanned{%
    \let\\=\origOr 
    \advance\last@column by\TabColSpan
    \decr\last@column
    \cell@wd=\z@
    \cell@kern=\z@
    \TMPcountA=\column
    \loop
	\advance\cell@wd by \ifcase\the\TMPcountA\TabColW@\fi\pt
    \ifnum\TMPcountA<\last@column
	\incr\TMPcountA
    \repeat
    \edef\TabJustHL{\ifcase\the\column\TabJustHS\else*\fi}%
    \ifx\TabJustHL\@default
	\let\TabJustHL=\TabJustHO
	\leftskip=\TabGutL 
	\ifx\TabJustHL\@both\else
	\ifx\TabJustHL\@left\else
	    \advance\leftskip by \fillskip
	\fi\fi
    \fi
    \edef\TabJustHR{\ifcase\the\last@column\TabJustHS\else*\fi}%
    \ifx\TabJustHR\@default
	\let\TabJustHR=\TabJustHO
	\rightskip=\TabGutR 
	\ifx\TabJustHR\@both\else
	\ifx\TabJustHR\@right\else
	    \advance\rightskip by \fillskip
	\fi\fi
    \fi
    \loosenup 
    \let\\=\relax
}

\def\cell@init@unspanned{%
    \csname cell@init@\number\column-\number\column\endcsname
}

\def\Endcell#1[#2]{%
	\ignorespaces #2%
	\ifhmode\unpenalty\par\fi 
	\ifnum\TabVSpan>1 \else
	  \TMPskipA=\TabGutB
	  \advance\TMPskipA by .5\rule@th
	  \ifdim \prevdepth>-1000pt \advance\TMPskipA by -\prevdepth \fi
	  \ifdim \TMPskipA<\z@
	    \advance\TMPskipA by\prevdepth
	  \fi
	  \ifx\TabJustV\@bot
	  \else
	    \advance\TMPskipA by 0pt plus 1fil
	    \relax
	  \fi
	  \vskip\TMPskipA 
	\fi
	\CellBotExec
    \egroup 
	\ifdim\cell@kern=\z@
           \wd\cell@box=\cell@wd
	   \box\cell@box 
	\else
	   \Endcell@negwd
	\fi
    \setbox\cell@box=\box\voidb@x	
    \column=\last@column
    \ignorespaces
}

\def\Endcell@negwd{%
    \wd\cell@box=\cell@wd
    \kern-\cell@kern\copy\cell@box\kern-\cell@kern
}

\def\Cell@II{%
    \hsize=\cell@wd
    \ifx\TabJustH\@default
    \else
      \cell@setHjustification
    \fi
    \parindent=\z@
    \parfillskip=\filskip
}

\def\cell@setHjustification{%
      \let\TabJustHL=\TabJustH
      \leftskip=\TabGutL 
      \ifx\TabJustHL\@both\else
      \ifx\TabJustHL\@left\else
	  \advance\leftskip by \fillskip
      \fi\fi
      \let\TabJustHR=\TabJustH \let\\=\or
      \rightskip=\TabGutR 
      \ifx\TabJustHR\@both\else
      \ifx\TabJustHR\@right\else
	  \advance\rightskip by \fillskip
      \fi\fi
}

\def\Endcell@ifsplit{%
    \Endcell@split
    \ifdim\cell@kern=\z@
       \hbox to\cell@wd{\copy0}%
       \setbox\xrow@boxsofar=\hbox{\unhbox\xrow@boxsofar
	 \hbox to\cell@wd{\copy\cell@box}}%
    \else
       \hbox to-\cell@wd{\kern-\cell@kern\copy0 \kern-\cell@kern}%
       \setbox\xrow@boxsofar=\hbox{\unhbox\xrow@boxsofar
	 \hbox to-\cell@wd{\kern-\cell@kern\copy\cell@box\kern-\cell@kern}}%
    \fi
}

\def\Endcell@split{%
    \let\next=\relax
    \TMPcountB=\vbadness \vbadness=10000
    \setbox2=\copy\cell@box		
    \splittopskip=\z@ \splitmaxdepth=\z@
    \setbox0=\vsplit\cell@box to\Tab@window
    \setbox0=\vbox{\unvbox0}
    \resulttrue
    \ifdim\ht\cell@box>\splittopskip\else \resultfalse\fi
    \ifdim\ht0=\z@ \resultfalse\fi
    \ifresult 
      \edef\CsA{\splitbottabheadmark}%
      \ifx\CsA\empty			
	\TMPdimenA=\Tab@window
	\advance\TMPdimenA by-\TabGutB \advance\TMPdimenA by-.5\rule@th
	\TMPdimenB=\TabGutT \advance\TMPdimenB by.5\rule@th
	\splittopskip=\TMPdimenB \splitmaxdepth=\maxdepth
	\setbox\cell@box=\box2 \setbox0=\vsplit\cell@box to\TMPdimenA
	\setbox0=\vtopbox to\TMPdimenA{%
	    \TMPdimenA=\dp0 \unvbox0 \prevdepth=\TMPdimenA
	    \TMPdimenA=\TabGutB \advance\TMPdimenA by.5\rule@th
	    \lineskiplimit=\z@ \baselineskip=\TMPdimenA
	    \hbox to\cell@wd{\hss}%
	    \ifx\TabJustV\@bot \else\vfill\fi
	}%
	\TMPdimenA=\ht\cell@box \advance\TMPdimenA by\dp\cell@box
	\ht\cell@box=\z@ \dp\cell@box=\TMPdimenA
      \else				
	\setbox2=\box\voidb@x		
	\setbox0=\vtopbox{%
	    \unvbox0
	    \global\setbox\brk@rule=\box\voidb@x
	    \ifduplicatinglastrule
	      \ifdim\lastkern=-.5\rule@th
		\unkern
		\global\setbox\brk@rule=\lastbox 
		\nointerlineskip\copy\brk@rule 
		\kern -.5\rule@th
		\ifdim\ht\brk@rule>-\dp\brk@rule
		  \global\setbox\brk@rule=\box\voidb@x
		\fi
	      \fi
	    \fi
	}%
	\setbox\cell@box=\vtop{%
	    \ifvoid\brk@rule \else	
		\kern .5\rule@th
		\nointerlineskip\box\brk@rule
		\kern-.5\rule@th
	    \fi
	    \unvbox\cell@box
	}%
      \fi
    \else				
	\ifdim\cell@ht=\@sentinel \else \ht0=\z@ \fi 
	\ifdim\ht0=\z@			
	    \setbox\cell@box=\box2 
	    \setbox0=\hbox to\cell@wd{\hss}%
	\else				
	    \setbox0=\box2 
	    \setbox\cell@box=\hbox to\cell@wd{\hss}%
	\fi
    \fi
    \vbadness=\TMPcountB		
    \next
}

\def\table@fixcellheights #1{%
  \setbox#1=\hbox{%
    \cell@ht=\ht#1\relax
    \unhbox#1{\table@fixnode}}%
}

\def\table@fixnode{%
  \setbox0=\lastbox
  \ifvoid0
    \TMPdimenA=\lastkern \unkern
    \ifdim\TMPdimenA=\z@
      \TMPskipA=\lastskip \unskip
      \ifdim\TMPskipA=\z@
      \else
	{\table@fixnode}%
	\hskip\TMPskipA
      \fi
    \else
      {\table@fixnode}%
      \kern\TMPdimenA
    \fi
  \else
    \ifvbox0
      {\table@fixnode}%
      \TMPdimenA=\wd0
      \setbox0=\vbox to\cell@ht{\unvbox0}%
      \wd0=\TMPdimenA
      \box0
    \else
      {\table@fixnode}%
      \unhbox0
    \fi
  \fi
}

\NEW\newif\ifintableheading
\NEW\newif\ifusingheading

\def\Tabheading#1[#2]{%
    \ignorespaces #2%
    \put@row\widowrowpenalty
    \incr\tabhead@number
    \global\setbox\ifodd\tabhead@number\tabheadboxA\else\tabheadboxB\fi=%
      \box\voidb@x
    \tabheadmark{\the\tabhead@number}%
    \incr\tabhead@number
    \TMPcountA=\ifodd\tabhead@number\tabheadboxA\else\tabheadboxB\fi
    \global\setbox\TMPcountA=\vbox
    \bgroup
	\needmigrationtrue
	\Tab@window=\@sentinel 
	\intableheadingtrue
        \TableChd@fonts
}

\def\Endtabheading#1[#2]{%
        \par #2%
	\put@row\widowrowpenalty
	\xdef\GcsA{\the\aftertabheadtoks}
    \egroup\GcsA
    \nointerlineskip
    \ifdim\Tab@window=\@sentinel
    \else
	\advance\Tab@window by-\ht\TMPcountA
	\advance\Tab@window by-\dp\TMPcountA
    \fi
    \copy\TMPcountA
    \tabheadmark{\the\tabhead@number}%
    \row=\z@
    \usingheadingtrue
    \duplicatinglastrulefalse
}

\def\TableChd@fonts{%
    \Tab@ord@font \TableChd@fnt \ea\let\ea\CsA\the\font \let\Tab@chd@font=\CsA
    \nullfont \tracinglostchars=1 
    \spaceskip=0pt 
    \savebaselineskip\offinterlineskip
}

\def\TableChd@fnt{\relax
  \ifx \alphatypesize\undefined \else
    \let\otherfontfam=\TableChdfntFfmOtherName
    \TableChdfntFfm
    \settypesize{\TableChdfntTsz}{\TableChdfntTszPercentmain}%
      {\TableChdfntTszIncrmain}{\TableChdfntTszOtherSize}%
    \TableChdfntTst
  \fi
}

\catcode`\@=11
\ifx\getcolfont\undefined \let\getcolfont=\relax \fi
\def\@dalign{d}

\def\TabColAlign{}
\def\initalignLandR{\def\alignL{\else\z@}\def\alignR{\else\z@}}

\def\initdecimalalign{%
  \ifx\TabColAlignRel\empty \initalignLandR
  \else \getalignpoints \def\TabColAlignRel{}
  \fi}

\let\origOr=\or

\def\getalignpoints{%
  \let\svdblblash=\\\let\\=\origOr\let\or=0%
  \def\alignL{}\def\alignR{}
  \column=0 \loop \ifnum\column<\TabColC \incr\column \addalignLandR \repeat
  \let\or=\origOr \let\\=\svdblblash }

\def\addalignLandR{
    \edef\TMPcsA{\ifcase\the\column\TabJustHS\else\TabJustHO\fi}
    \ifx\TMPcsA\@default \edef\TMPcsA{\TabJustHO}\fi 
    \ifx\TMPcsA\@both    \edef\TMPcsA{\@center}\fi   
    \TMPdimenA=\ifcase\the\column\TabColW@\fi
    \setbox\z@=\hbox
      {\Tab@ord@font\getcolfont\hskip\TabGutL\hskip\TabGutR}
    \advance\TMPdimenA-\wd\z@ \TMPdimenB=\z@
    \ifx\TMPcsA\@center
      \divide\TMPdimenA by2 \TMPdimenB=\TMPdimenA
    \else\ifx\TMPcsA\@left
      \TMPdimenB=\TMPdimenA
      \TMPdimenA=\z@
    \else\ifx\TMPcsA\@right
    \else\ifx\TMPcsA\@dalign
      \TMPdimenB=\TMPdimenA
      \TMPdimenA=\expandafter
        \ifcase\the\column\TabColAlignRel\else 0.5\fi\TMPdimenA
      \advance\TMPdimenB by-\TMPdimenA
    \fi\fi\fi\fi
    \edef\alignL{\alignL\or\the\TMPdimenA\space}%
    \edef\alignR{\alignR\or\the\TMPdimenB\space}%
}

\def\tabaligncase@{%
      \afterassignment\TMPtoksB
      \TMPtoksA=\expandafter\ifcase\the\column\TabColAlign\fi{}{}%
      \ifx\TMPcsA\@dalign
        \tabaligncase@dalign
      \else\ifx\TMPcsA\@left
        \tabaligncase@left
      \else\ifx\TMPcsA\@right
        \tabaligncase@right
      \fi\fi\fi
}
\def\tabaligncase@dalign{%
        \setbox\TMPboxA=\hbox{\let\or=\origOr
		\Tab@ord@font\getcolfont\the\TMPtoksA}%
        \setbox\TMPboxB=\hbox{\let\or=\origOr
		\Tab@ord@font\getcolfont\the\TMPtoksB}%
        \advance\TMPdimenA-\wd\TMPboxA
        \advance\TMPdimenA-\wd\TMPboxB
        \divide\TMPdimenA by2 \TMPdimenB=\TMPdimenA
        \advance\TMPdimenA by\wd\TMPboxA
        \advance\TMPdimenB by\wd\TMPboxB
}

\def\tabaligncase@left{%
	\setbox\TMPboxA=\hbox{\let\or=\origOr
		\Tab@ord@font\getcolfont\the\TMPtoksA}%
        \TMPdimenB=\TMPdimenA
        \TMPdimenA=\wd\TMPboxA
        \advance\TMPdimenB by-\TMPdimenA
}

\def\tabaligncase@right{%
        \setbox\TMPboxB=\hbox{\let\or=\origOr
		\Tab@ord@font\getcolfont\the\TMPtoksB}%
        \TMPdimenB=\wd\TMPboxB
        \advance\TMPdimenA by-\TMPdimenB
}

\def\TabAlign #1#2{
  \ifvmode\leavevmode\else \unpenalty\unskip\hskip\parfillskip\break\fi
  \TMPdimenA=\ifcase\the\column\alignL\fi
  \setbox\TMPboxA=\hbox{\avoidvmode #1}%
  \iftabalignOK
    \advance\TMPdimenA by-\wd\TMPboxA
    \vadjust{}\kern\TMPdimenA\unhbox\TMPboxA
    \TMPdimenB=\ifcase\the\column\alignR\fi
  \else
    #1%
  \fi
  \setbox\TMPboxB=\hbox{\avoidvmode #2}%
  \iftabalignOK
    \advance\TMPdimenB by-\wd\TMPboxB
    \unhbox\TMPboxB\unskip\kern\TMPdimenB
    \break
  \else
    #2%
  \fi
  \ignorespaces 
}

\def\avoidvmode{%
  \global\tabalignOKtrue
  \let\DispEqn=\avoidDispEqn
  \let\InlEqn=\avoidInlEqn
  \let\LstList=\avoidLstList
}

\def\avoidDispEqn[#1]#2\EndDispEqn[]{%
  \PerPgmessage{equation in table can't be character-aligned}%
  \global\tabalignOKfalse
}

\def\avoidInlEqn[#1]#2\EndInlEqn[]{%
  \PerPgmessage{equation in table can't be character-aligned}%
  \global\tabalignOKfalse
}

\def\avoidLstList[#1]#2\LstEndlist[]{%
  \PerPgmessage{list in table can't be character-aligned}%
  \global\tabalignOKfalse
}

\newif\iftabalignOK

\ifx\undefined\tableoutput
  \newoutput{\tableoutput}
\fi

\def\tableoutput{%
	\PubTeXuserstatus=1
	\edef\bottabheadmark{\bottabheadmark}
	\ifx\bottabheadmark\empty
	  {\lineskiplimit=\z@ \the\output} 
	\else
	  \brokentableoutput
	\fi
}

\def\brokentableoutput{%
  {\setbox0=\vbox to12pt{}\nointerlineskip\box0\kern-12pt}%
  \resultfalse 
  \TMPcountA=\bottabheadmark\relax
  \ifnum\TMPcountA>0
    \incr\TMPcountA
    \ifnum\TMPcountA<\tabhead@number\else\resulttrue \fi 
  \fi
  \global\setbox\brk@rule=\box\voidb@x
  \ifduplicatinglastrule
    \ifresult\else 
        \setbox255=\vbox{%
	  \unvbox\@cclv 
          \TMPdimenA=\lastkern 
          \unkern
	  \global\setbox\brk@rule=\lastbox 
          \setbox\TMPboxA=\copy\brk@rule
	  \ifdim\ht\brk@rule>-\dp\brk@rule
	    \global\setbox\brk@rule=\box\voidb@x
	  \fi
          \nointerlineskip
          \box\TMPboxA
          \ifdim\TMPdimenA=0pt \else\kern\TMPdimenA\fi
        }%
    \fi
  \fi
  {\lineskiplimit=\z@ \the\output} 
  \ifresult \copy\ifodd\TMPcountA\tabheadboxB\else\tabheadboxA\fi \fi
  \ifvoid\brk@rule \else	
    \nointerlineskip\box\brk@rule 
  \fi
  \global\donewwindowtrue
}

	\def\averagewordsize{6 }
	\def\averageinterhyphen{3 }
\def\loosenup{
    \TMPdimenA=.425em 
    \ifdim\TMPdimenA>\z@
        \loosenup@
    \else
        \spaceskip=\z@\xspaceskip=\z@\relax
    \fi
}

\def\loosenup@{%
    \TMPdimenA=.425em 
    \TMPcountA=\hsize 
    \advance\TMPcountA-\leftskip \advance\TMPcountA-\rightskip
    \divide\TMPcountA\TMPdimenA 
    \divide\TMPcountA\averagewordsize 
    \decr\TMPcountA 
    \multiply\TMPdimenA\averageinterhyphen 
    \ifnum\TMPcountA>1
      \divide\TMPdimenA\TMPcountA \fi 
    \TMPdimenA=.7\TMPdimenA
    \advance\TMPdimenA-\fontdimen4\font \divide\TMPdimenA by2
    \ifdim\TMPdimenA>\fontdimen3\font 
	\spaceskip=\fontdimen2\font plus\TMPdimenA minus\fontdimen4\font
        \xspaceskip=\spaceskip 
        \advance\xspaceskip by\fontdimen7\font 
    \else 
	\spaceskip=\z@\xspaceskip=\z@\relax
    \fi
    \ifnum\tolerance<800 \tolerance=800 \fi
}

\def\Bl{\leavevmode\vrule width0pt\kern\fontdimen2\font\nobreak}

\ifx\undefined\outeradjust
  \def\outeradjust#1{}
  \let\needmigrationtrue=\relax
  \def\adjustinto{\chardef\@adjustinto}
\fi

\ifx\undefined\RVBdropmigrations
  \def\RVBdropmigrations#1{}%
  \def\RVBadjustinto{\chardef\@adjustinto}
\fi

\ifx\notableshading\undefined \else \endinput \fi

\def\CellShade{0}
\def\CellFreq{0}

\NEW\newif\ifcellgrayused
\NEW\newbox\shade@box@tl 
\NEW\newbox\shade@box@br 
\NEW\newcount\shade@num  
\chardef\shade@num@max=40

\def\newshade@num{%
  \global\incr\shade@num
  \ifnum\shade@num>\shade@num@max
    \global\shade@num=1 \fi
}

\def\RowLeftExec{%
    \global\cellgrayusedfalse
    \global\setbox\shade@box@tl=\hbox{\kern\Tab@leftskip}%
    \global\setbox\shade@box@br=\hbox{\kern\Tab@leftskip}%
}

\def\CellBotExec{%
  \relax
  \ifnum\CellShade=0
    \global\setbox\shade@box@tl=\hbox{\unhbox\shade@box@tl\kern\cell@wd}%
    \global\setbox\shade@box@br=\hbox{\unhbox\shade@box@br\kern\cell@wd}%
  \else
    \global\cellgrayusedtrue
    \global\setbox\shade@box@tl=\hbox{%
      \unhbox\shade@box@tl
      \newshade@num
      \shadepointspecial\shade@num
      \kern\cell@wd}%
    \global\setbox\shade@box@br=\hbox{%
      \unhbox\shade@box@br
      \kern\cell@wd
      \TMPcountA=\shade@num 
      \newshade@num
      \shadepointspecial\shade@num
      \edef\CellGray{\shadetogray\CellShade}%
      \shadeboxspecial\TMPcountA\shade@num\CellGray\CellFreq}%
  \fi
}

\def\putrowExec{%
  \ifcellgrayused 
    \nointerlineskip
    \TMPdimenA=\ht\row@box
    \advance\TMPdimenA\dp\row@box
    \vbox to \z@{%
      \shadedictspecial  
      \shadebeginspecial 
      \nointerlineskip
      \box\shade@box@tl  
      \kern \TMPdimenA
      \nointerlineskip
      \box\shade@box@br	 
      \kern-\TMPdimenA
      \shadeendspecial   
    }%
    \nointerlineskip
    \copy\lastrrule@box 
  \fi
}

\ifx\undefined\specialbop \let\specialbop=\special \fi

\def\GrayDictSpecial{%
}

\def\GrayPointSpecial #1{%
  \special{ps::\number#1 GRSP}%
}

\def\GrayBoxSpecial #1#2#3#4{
  \special{ps::\number#1 \space \number#2 \space #3 \number#4 \space GRFB}%
}

\def\GrayBeginSpecial{%
  \special{ps::save}%
}

\def\GrayEndSpecial{%
  \special{ps::restore}%
}

\def\shadepointspecial	{\GrayPointSpecial}
\def\shadeboxspecial	{\GrayBoxSpecial}
\def\shadedictspecial	{\GrayDictSpecial}
\def\shadebeginspecial	{\GrayBeginSpecial}
\def\shadeendspecial	{\GrayEndSpecial}

\def\shadetogray #1{
  \ifcase#1
    1\space
    \or\grayA\space
    \or\grayB\space
    \or\grayC\space
    \or\grayD\space
    \or\grayE\space
    \else
    0\space
  \fi
}

\ifx\undefined\grayA \def\grayA{.99}\fi
\ifx\undefined\grayB \def\grayB{.95}\fi
\ifx\undefined\grayC \def\grayC{.90}\fi
\ifx\undefined\grayD \def\grayD{.80}\fi
\ifx\undefined\grayE \def\grayE{.75}\fi

\def\SGMPcomment#1{}

\def\SGMPdefifnonempty#1#2{\def\tempcs{#2}%
	\ifx \tempcs\empty \else
		\def#1{#2}%
	\fi
}

\let\mdef=\SGMPdefifnonempty

\newif\ifSGMPkeeplooping

\newtoks\SGMPttlpgtoks	\SGMPttlpgtoks={}

\long\def\SGMPttlpgmisc#1{\relax
	\expandafter\SGMPttlpgtoks\expandafter{\the\SGMPttlpgtoks
				\vskip6pt plus6pt minus4pt#1\par}%
}

\def\SGMPpreface{%
	\edef\pgn{\the\count0}\begin{titlepage}
	\pagestyle{empty} \vglue1.0in\leftline{\huge Preface}
	\vskip4pc\normalsize\noindent
}

\def\SGMPendpreface{%
	\end{titlepage}\global\advance\count0 by\pgn
	\global\advance\count0 by 1
}

\catcode`\@=11
\def\SGMPmaketitle{%
	\gdef\@thanks{\begin{center}\normalsize\the\SGMPttlpgtoks\end{center}}%
	\maketitle
}

\def\SGMPref#1{%
	\@ifundefined{r@x:#1}{\pageref{pg:#1}}{\ref{x:#1}}%
}
\catcode`\@=12

\def\SGMPunident{}
\def\SGMPlab#1{\ifx#1\SGMPunident\else\label{x:#1}\fi}

\newtoks\SGMPstackA		\global\SGMPstackA={}
\newtoks\SGMPstackB		\global\SGMPstackB={}

\def\push #1\onto#2{\relax
  \global\SGMPstackA=\expandafter{#1}\global\SGMPstackB=\expandafter{#2}%
  \xdef#2{{\the\SGMPstackA}{\the\SGMPstackB}}%
}

\def\gsetSGMPstackB{\global\SGMPstackB}

\def\pop #1\from#2{\relax
  \afterassignment\gsetSGMPstackB \global\SGMPstackA=#2{}{}%
  \xdef#1{\the\SGMPstackA}\xdef#2{\the\SGMPstackB}%
}

\def\SGMPbeginList#1#2{%
	\def\SGMPlisttype{#1}%
	\push \SGMPendList\onto \SGMPendListstack
	\push \SGMPitem\onto \SGMPitemstack
	\ifx \SGMPlisttype\SGMPlisttext
		\def\SGMPendList{%
			\end{description}%
			\pop \SGMPendList\from \SGMPendListstack
			\pop \SGMPitem\from \SGMPitemstack
		}%
		\def\next{\begin{description}}%
		\def\SGMPitem{\item[#2]}%
	\else \ifx \SGMPlisttype\SGMPlistnone
		\def\SGMPendList{%
			\end{description}%
			\pop \SGMPendList\from \SGMPendListstack
			\pop \SGMPitem\from \SGMPitemstack
		}%
		\def\next{\begin{description}}%
		\def\SGMPitem{\item[]}%
	\else \ifx \SGMPlisttype\SGMPlistbulleted
		\def\SGMPendList{%
			\end{itemize}%
			\pop \SGMPendList\from \SGMPendListstack
			\pop \SGMPitem\from \SGMPitemstack
		}%
		\def\next{\begin{itemize}}%
		\def\SGMPitem{\item}%
	\else \ifx \SGMPlisttype\SGMPlistsquare
		\def\SGMPendList{%
			\end{itemize}%
			\pop \SGMPendList\from \SGMPendListstack
			\pop \SGMPitem\from \SGMPitemstack
		}%
		\def\next{\begin{itemize}}%
		\def\SGMPitem{\item[$\Box$]}%
	\else
		\def\SGMPendList{%
			\end{enumerate}%
			\pop \SGMPendList\from \SGMPendListstack
			\pop \SGMPitem\from \SGMPitemstack
		}%
		\def\next{\begin{enumerate}}%
		\def\SGMPitem{\item}%
	\fi \fi \fi \fi
	\next
}

\let\SGMPendList=\relax		\let\SGMPitem=\relax
\let\SGMPendListstack=\relax	\let\SGMPitemstack=\relax
\def\SGMPlisttext{text} \def\SGMPlistnone{none}
\def\SGMPlistbulleted{bulleted} \def\SGMPlistsquare{square}

\def\SGMPcite#1#2{\cite{#1}}

\makeindex
\def\SGMPindex#1#2{\def\tempcs{#1}%
	\ifx \tempcs\SGMPvalueyes
		#2%
	\fi
	\index{#2}%
}
\def\SGMPvalueyes{yes}
\begingroup \catcode`\@=11
\newread\dfe@ \gdef\dfe#1#2#3{\relax
       \immediate\openin\dfe@=#1 \ifeof\dfe@#3\else#2\fi
       \immediate\closein\dfe@}
\gdef\SGMPstartindex{\relax\ifx\@indexfile\undefined\else
       \closeout\@indexfile \fi\begin{theindex}
       \def\indexentry##1##2{\item##1 ##2}}
\gdef\SGMPfinishindex{\dfe{\jobname.ind}{\def\next{\input
       \jobname.ind}}{\let\next=\relax}\ifx\next\relax \dfe
       {\jobname.idx}{\def\next{\input \jobname.idx}}{\relax}\fi
       \next \end{theindex}}
\endgroup

\def\aalign#1{\leavevmode\vbox{\baselineskip=0pt \lineskiplimit.25ex
  \ialign{##\crcr#1\crcr}}}
\def\SGMPring#1{\aalign{\hidewidth\char"17\hidewidth\cr\noalign{\kern-1.2ex}#1}}

\let\SGMPnewline=\\

\newtoks\TexMacPairEndtextoks

\def\stack#1#2{{#1\atop #2}}

\def\SGMPgobble#1{}

\def\SGMPlim#1{\def\tempcs{#1}%
	\ifx \tempcs\empty
		\let\SGMPdolim=\displaylimits
	\else \if #1c
		\let\SGMPdolim=\limits
	\else \if #1r
		\let\SGMPdolim=\nolimits
	\else
		\let\SGMPdolim=\relax
	\fi \fi \fi
}

\def\Rad#1{%
	\begingroup
	\def\RadTempCs{{#1}}\let\RdxTempCs=\empty
}

\def\DoRad{%
	\relax
	\ifx \RdxTempCs\empty
		\sqrt\RadTempCs
	\else
		\root \RdxTempCs \of \RadTempCs
	\fi
	\endgroup
}

\def\LeftPost#1{\csname LP#1\endcsname}
\def\RightPost#1{\csname RP#1\endcsname}

\def\getchar #1#2\endgetchar{\def\gotchar{#1}\def\ungotchars{#2}}

\def\SGMPmathgrk#1{%
    \def\ungotchars{#1}%
    \SGMPkeeploopingtrue
    \loop
	\expandafter\getchar\ungotchars\endgetchar
	\ifx \gotchar\empty \def\gotchar{0}\fi
	\count255=\expandafter`\gotchar\relax
	\advance\count255 by -49
	\ifcase \count255
		\nabla
	\or	\varpi
	\or	\varepsilon
	\or	\varphi
	\or		
	\or	\partial
	\or\or
	\or	\varrho
	\or\or\or\or\or\or\or
	\or	A%
	\or	B%
	\or	X%
	\or	\Delta
	\or	E%
	\or	\Phi
	\or	\Gamma
	\or	H%
	\or	I%
	\or		
	\or	K%
	\or	\Lambda
	\or	M%
	\or	N%
	\or	O%
	\or	\Pi
	\or	\Theta
	\or	P
	\or	\Sigma
	\or	T%
	\or	\Upsilon
	\or
	\or	\Omega
	\or	\Xi
	\or	\Psi
	\or	Z%
	\or\or\or\or\or\or
	\or	\alpha
	\or	\beta
	\or	\chi
	\or	\delta
	\or	\epsilon
	\or	\phi
	\or	\gamma
	\or	\eta
	\or	\iota
	\or	\vartheta
	\or	\kappa
	\or	\lambda
	\or	\mu
	\or	\nu
	\or	o%
	\or	\pi
	\or	\theta
	\or	\rho
	\or	\sigma
	\or	\tau
	\or	\upsilon
	\or	\varsigma
	\or	\omega
	\or	\xi
	\or	\psi
	\or	\zeta
	\else
	\fi
	\relax
	\ifx \ungotchars\empty \SGMPkeeploopingfalse \fi
	\ifSGMPkeeplooping
    \repeat
}

\catcode`\@=11
\def\eqalign#1{\null\,\vcenter{\openup\jot\m@th
  \ialign{\strut\hfil$\displaystyle{##}$&$\displaystyle{{}##}$\hfil
      \crcr#1\crcr}}\,}
\catcode`\@=12

\def\MthAcnt#1#2{#2{#1}}

\def\SGMPgraphic#1#2#3#4#5#6#7{{%
	\def\type{#3}
	\def\imresdefault{#4}
	\def\imresvalue{#5}
	\def\picresdefault{#6}
	\def\picresvalue{#7}

	\def\yes{yes}
	\def\drawing{drawing}
	\def\image{image}
	\def\height{4in}

	\ifx\type\drawing	
	    \vbox to\height{%
			\special{pub: pubdraw #2 #10}
			\vfil}
	\else\ifx\type\image	
	    \ifx\imresdefault\yes
		\vbox to\height{%
			\vfil
			\special{pub: sunbitmap #2 #10 0}}
	    \else
		\vbox to\height{%
			\vfil
			\special{pub: sunbitmap #2 #10 \imresvalue}}
	    \fi
	\else			
	    \ifx\picresdefault\yes
		\vbox to\height{%
			\vfil
			\special{pub: sunbitmap #2 #10 0}}
	    \else
		\vbox to\height{%
			\vfil
			\special{pub: sunbitmap #2 #10 \picresvalue}}
	    \fi
	\fi\fi
}}

\ifx\TMPcountA\undefined
  \catcode`\!=10
\else
  \catcode`\!=14
\fi
!\newcount\TMPcountA
!\newcount\TMPcountB
!\newdimen\TMPdimenA
!\newdimen\TMPdimenB
\catcode`\!=12

\def\SGMPTabcnvtlist#1#2{%
	\def\tempcsA{#1}%
	\def#2{}%
	\ifx \tempcsA\empty \else
	    \SGMPkeeploopingtrue
	    \loop
		\expandafter\SGMPparsetablelist\tempcsA:::\endSGMPparsetablelist#2
		\ifx \tempcsA\empty
			\SGMPkeeploopingfalse
		\fi
		\ifSGMPkeeplooping
	    \repeat
	\fi
}
\def\SGMPparsetablelist #1:#2::#3\endSGMPparsetablelist#4{%
	\def\tempcsA{#2}%
	\expandafter\def\expandafter#4\expandafter{#4\\#1}%
}

\def\SGMPTabColW#1{\SGMPTabcnvtlist{#1}\TabColW}

\def\SGMPTableWd#1{\def\tempcs{#1}%
	\ifx \tempcs\SGMPabs
		\def\TableWd{A}%
	\else
		\def\TableWd{R}%
		\def\TableWdRPct{#1}%
	\fi
}

\def\SGMPabs{abs}

\def\SGMPbeginTable#1#2#3#4#5#6#7#8#9{%
	\edef\TabRuleVO{\TabRuleVI}\edef\TabRuleHO{\TabRuleHI}%
	\SingleRuleWidthInPixels=6
	\Table[\SGMPTableWd{#9}\mdef\TableJust{#6}\SGMPTabColW{#5}%
		\mdef\TabJustVO{#8}\mdef\TabJustVH{#8}%
		\mdef\TabRuleHI{#7}\mdef\TabRuleHO{#7}\mdef\TabRuleHH{#7}%
		\mdef\TabRuleVI{#3}\mdef\TabRuleVO{#3}%
		\mdef\TabJustHO{#1}\mdef\TabJustHH{#1}%
		\SGMPTabJustHS{#4}%
		\SGMPTabRuleVS{#2}]%
	\let\SGMPnewline=\newline
}

\def\newline{\relax
	\ifvmode
		\vskip\baselineskip
	\else
		\unskip\vadjust{}\nobreak\hfil\break\vadjust{}\ignorespaces
	\fi
}

\def\SGMPTabJustHS#1{\SGMPTabcnvtlist{#1}\TabJustHS}

\def\SGMPTabRuleVS#1{\SGMPTabcnvtlist{#1}\TabRuleVS

\expandafter\SGMPrminitialdblsh\TabRuleVS\\\\\\\endSGMPrminitialdblsh\TabRuleVS
}

\def\SGMPrminitialdblsh\\#1\\\\#2\endSGMPrminitialdblsh#3{\def#3{#1}}

\def\SGMPTabRuleHS#1{\SGMPTabcnvtlist{#1}\TabRuleHS}

\begin{document}
\parskip = 5pt
\title{New model for crack growth using random walkers
       \footnote{HLRZ preprint 106/92}}
\author{Peter Ossadnik\\
H\"ochstleistungsrechenzentrum (HLRZ)\\
Forschungszentrum J\"ulich GmbH\\
Postfach 1913, W-5170 J\"ulich, Germany}
\maketitle
\begin{abstract}
In close analogy to diffusion limited aggregation (DLA) and
inspired by a work of Roux, a random walker algorithm is
constructed to solve the problem of crack growth in an elastic medium.
In contrast to conventional lattice approaches, the stress field is not
calculated throughout the whole medium, but random walkers are used
to detect only the hot sites on the surface of the crack. There, an
analytically calculated Green{'}s function is used to determine the
stress field. The complicated boundary condition on the
crack surface is simulated by a special sticking-rule walk.
Using this new method we
generate crack-clusters up to sizes of 20,000 particles on simple
workstations within reasonable time.
We simulate several different boundary conditions,
like uniaxial tensile, pure shear and isotropic tensile load.
We study the influence of several parameters representing the material
strength or fatigue. Furthermore we study the effect of anisotropic
walks. As a result, we reproduce
with this new model typical experimental crack shapes and are able
to simulate the essential features of realistic cracks.
\end{abstract}

\def\XRefId{}\section{\SGMPlab\XRefId Introduction}

\par The study of non-equilibrium growth models has been extremely
popular in the past few years. Especially the Laplacian growth
phenomenon has been studied extensively and applications of it can be
found in a large variety of fields, e. g. electrodeposition
\SGMPcite{brady_ball_nature}{}, fluid-fluid displacement
\SGMPcite{daccord_nittmann_stanley}{} or growth of
bacteria colonies \SGMPcite{matsushita_fujikawa}{}. One
reason for its popularity is, that a numerical simulation of this
process can be done extremely efficiently using diffusion-limited
aggregation (DLA) \SGMPcite{witten_sander}{}. Using
highly optimized programs \SGMPcite{tolman_meakin,ossadnik_physica_a_1991}{} it
is possible
to grow huge DLA clusters containing up to 50 million particles
\SGMPcite{ossadnik_30m,mandelbrot_hamburg_proceedings}{}.\par

\par However, this paper is not devoted to DLA, but inspired by this
efficiency and success of DLA and by a work of Roux
\SGMPcite{roux_random_walker}{} we will attack the problem of
crack growth in an elastic medium
\SGMPcite{fracture_of_disordered_media,fractals_and_disordered_systems,disorder_and_fracture}{}.
Here, to an elastic material a certain load is applied and eventually
cracks develop, which grow until the material breaks apart. Although
highly developed numerical techniques, which are mainly based on lattice
models, exist \SGMPcite{herrmann_physica_a_1990,hinrichsen_hansen_roux}{}, the
simulation of crack growth is even on today{'}s supercomputers restricted
to a few thousand broken bonds or even less. Thus, new techniques are
urgently needed, which might allow for large crack growth simulations
maybe on workstations.\par

\par As was pointed out by many researchers
\SGMPcite{kertesz_in_disorder_and_fracture,roux_in_statistical_models}{},
there exist some remarkable formal similarities between the vectorial
elasticity problem {---} which is governed by the Lam\'e equation
{---} and the scalar electricity problem {---} which is governed by
the Laplace equation. It was even shown by Roux, that it is formally
possible to construct a Green{'}s function of the Lam\'e equation
using a certain type of vectorial random walkers in a similar way one
constructs a Green{'}s function of the Laplace equation. The latter one
can of course simply be interpreted as a density of random walkers and
is thus much easier to implement than a formal, vectorial random walker,
which is much more difficult to interpret. However, although Roux{'}s
ideas are very clear, an actual numerical implementation of them and a
check of their practicability is still missing.\par

\par In the current paper we will discuss his ideas and will present a
new algorithm for crack growth. For this purpose we will construct a
DLA-like random walker method, in which the random walkers are used to
find the {``}hot{''} (highly stressed) sites along the surface of
a crack, i.\space e.\space those
which are most likely to break. Here, {`}DLA-like{'} is not meant in the
sense of {`}probabilistic{'} used in
\SGMPcite{louis_guinea,meakin_li_sander_louis_guinea,meakin_li_sander_yan_guinea_pla_louis,louis_guinea_flores}{}
to characterize these crack growth models, but in a sense that we
explicitly use random walkers to transport information from the
boundaries towards a crack.\par

\par The use of random walkers will introduce disorder and noise into
the simulation. But since real material are typically spatially
disordered and inhomogeneous, this noise can be justified and
interpreted physically as an inherit randomness. It does not have to be
introduced by hand like in typical lattice models, but is an implicit
feature of the model.\par

\par The outline of this paper is as follows. In section \SGMPref{Roux} we
will give a short introduction into the problem of crack
growth. We will briefly present and extend Roux{'}s ideas and we will
discuss the reasons, why a direct numerical simulation of his method is
not practical. In section \SGMPref{Algorithm} we will introduce a new
model for crack growth by drawing an analogy to DLA. We will use random
walkers and an analytically calculated Green{'}s function to transport the
information about surface forces towards a crack, which is represented
as a cluster of particles. We will also discuss the treatment of the
boundary condition along the crack, which turns out to be the most
difficult ingredient of the model. Then, in section \SGMPref{Results},
we shall present the results of this model and will study several
variations of it in order to understand the influence of the numerical
parameters. Here, we will compare our model to experimental and
analytical results. In section \SGMPref{Discussion} we will critically
discuss the method and, finally, in section \SGMPref{Summary} we will
summarize our results.\par

\def\XRefId{Roux}\section{\SGMPlab\XRefId Relation between crack growth and
random walks}

\par Consider a
\(
d\)
 dimensional infinite elastic medium with Poisson
ratio
\(
\SGMPmathgrk{n}\)
. In equilibrium this system can be described in terms
of the stress field
\(
\underline{\underline{\SGMPmathgrk{s}}}\)
 which obeys
\def\XRefId{equilibrium}
\begin{equation}\SGMPlab\XRefId\vcenter{\halign{\strut\hfil#\hfil&#\hfil\cr
$\displaystyle{\hbox{\rm Div}\underline{\underline{\SGMPmathgrk{s}}}=0.}$\cr
}}\end{equation}

Together with Hooke{'}s law this can be reformulated in terms of the
displacement field
\(
\MthAcnt {u}{\matrix{}}\)
 and one obtains the Lam\'e equation
\def\XRefId{lame}
\begin{equation}\SGMPlab\XRefId\vcenter{\halign{\strut\hfil#\hfil&#\hfil\cr
$\displaystyle{{\left\LeftPost{par}1-\SGMPmathgrk{n}{\left
\LeftPost{par}d-1\right\RightPost{par}}
\right\RightPost{par}}\nabla ^{2}\MthAcnt {u}{\matrix{}}
+\nabla {\left\LeftPost{par}\nabla {\ifmmode\cdot\else\.\fi}\MthAcnt
{u}{\matrix{}}
\right\RightPost{par}}=0.}$\cr
}}\end{equation}

A crack in such a system can be interpreted as an additional force free
surface, which thus has to obey the boundary condition
\def\XRefId{boundary}
\begin{equation}\SGMPlab\XRefId\vcenter{\halign{\strut\hfil#\hfil&#\hfil\cr
$\displaystyle{\underline{\underline{\SGMPmathgrk{s}}}{\ifmmode\cdot\else\.\fi}\MthAcnt
{n}{\matrix{}}
=0.}$\cr
}}\end{equation}

Here one has to point out, that unlike in DLA this boundary condition at
the crack surface is not formulated in terms of the displacement field
{---} which corresponds to the potential in the Laplace equation
{---} but in terms of the stress field, which is related via Hooke{'}s
law to a spatial derivative of
\(
\MthAcnt {u}{\matrix{}}\)
. As we will show later, this peculiarity requires a
special treatment of the crack surfaces.\par

\par Apart from those equilibrium conditions one can formulate the
condition for the growth of a crack in terms of the stress component
parallel to the crack surface
\(
\SGMPmathgrk{s}_{\parallel }\)
. This has to exceed a critical value
\(
\SGMPmathgrk{s}_{c}\)
, which is directly connected to the intermolecular
cohesion forces, and then the growth velocity of the crack
\(
v_{n}\)
 is determined by
\def\XRefId{growth}
\begin{equation}\SGMPlab\XRefId\vcenter{\halign{\strut\hfil#\hfil&#\hfil\cr
$\displaystyle{v_{n}\propto {\left\LeftPost{par}\SGMPmathgrk{s}_{\parallel }
-\SGMPmathgrk{s}_{c}\right\RightPost{par}}^{\SGMPmathgrk{h}}
}$\cr
}}\end{equation}

where
\(
\SGMPmathgrk{h}\)
 is a heuristic parameter which is often simply set to
one. This rule for crack growth only takes into account crack growth
through cleavage and completely ignores bending terms. The equations
(\SGMPref{lame}, \SGMPref{boundary}, \SGMPref{growth}) formulate
the problem of crack growth as a nonlinear moving boundary
problem.\par
\vskip 5pt
\par To give a closed picture of the problem we shall now recall very
briefly the main ideas of Roux \SGMPcite{roux_random_walker}{} in order to be
able to extend and
comment them. The fundamental idea of his work is to define a Markovian
random walk process that constructs a matrix, which finally leads to a
Green{'}s function of the Lam\'e equation. This is done in complete
analogy to the Laplacian case, which we will find as a special case of
the following derivation.\par

\par Therefore, consider an elastic material with an unknown
displacement field
\(
\MthAcnt {u}{\matrix{}}\)
, which is a result of the boundary conditions. Our
task is to obtain this field at point
\(
\MthAcnt {x}{\matrix{}}\)
 in terms of the boundary displacements. Therefore
consider the following Markovian process: A random walker starts at
point
\(
\MthAcnt {x}{\matrix{}}\)
 and makes a first, elementary jump of length
\(
r\)
 into a random direction
\(
\MthAcnt {e}{\matrix{}}_{1}\)
. After averaging over all possible directions
\(
\MthAcnt {e}{\matrix{}}_{1}\)
 one defines at
\(
\MthAcnt {x}{\matrix{}}\)
 a new vector
\(
{\left\LeftPost{par}Q_{r}\MthAcnt {u}{\matrix{}}\right\RightPost{par}}
{\left\LeftPost{par}\MthAcnt {x}{\matrix{}}\right\RightPost{par}}
\)
 as
\def\XRefId{}
\begin{equation}\SGMPlab\XRefId\vcenter{\halign{\strut\hfil#\hfil&#\hfil\cr
$\displaystyle{{\left\LeftPost{par}Q_{r}\MthAcnt {u}{\matrix{}}
\right\RightPost{par}}{\left\LeftPost{par}\MthAcnt {x}{\matrix{}}
\right\RightPost{par}}={\left\LeftPost{ang}\underline{\underline{Q}}
{\left\LeftPost{par}
\MthAcnt {e}{\matrix{}}_{1}\right\RightPost{par}}
{\ifmmode\cdot\else\.\fi}\MthAcnt {u}{\matrix{}}{\left
\LeftPost{par}\MthAcnt {x}{\matrix{}}
+r\MthAcnt {e}{\matrix{}}_{1}\right\RightPost{par}}
\right\RightPost{ang}}_{\MthAcnt {e}{\matrix{}}_{%
1}}}$\cr
}}\end{equation}

where the elementary step matrix
\(
\underline{\underline{Q}}{\left\LeftPost{par}\MthAcnt {e}{\matrix{}}
\right\RightPost{par}}\)
 is defined as
\def\XRefId{}
\begin{equation}\SGMPlab\XRefId\vcenter{\halign{\strut\hfil#\hfil&#\hfil\cr
$\displaystyle{\underline{\underline{Q}}{\left\LeftPost{par}\MthAcnt
{e}{\matrix{}}
\right\RightPost{par}}_{ij}={\left\LeftPost{par}\SGMPmathgrk{a}
{\ifmmode\cdot\else\.\fi}
\underline{\underline{1}}+{\left\LeftPost{par}1-
\SGMPmathgrk{a}\right\RightPost{par}}
{\ifmmode\cdot\else\.\fi}d{\ifmmode\cdot\else\.\fi}\MthAcnt {e}
{\matrix{}}\otimes \MthAcnt {e}{\matrix{}}
\right\RightPost{par}}_{ij}\equiv {\left\LeftPost{par}\SGMPmathgrk{a}
{\ifmmode\cdot\else\.\fi}\SGMPmathgrk{d}_{ij}+{\left\LeftPost{par}1-
\SGMPmathgrk{a}\right\RightPost{par}}
{\ifmmode\cdot\else\.\fi}d{\ifmmode\cdot\else\.\fi}e_{i}e_{j}
\right\RightPost{par}}}$\cr
}}\end{equation}

and
\(
\SGMPmathgrk{a}\)
 is a parameter that will be determined later.\par

\par As usual in Markov processes this elementary step is iterated.
Consequently, after performing
\(
N\)
 independent steps {--} each of length
\(
r\)
 {--} one defines at
\(
\MthAcnt {x}{\matrix{}}\)
 the new vector
\def\XRefId{elementary_walk}
\begin{equation}\SGMPlab\XRefId\vcenter{\halign{\strut\hfil#\hfil&#\hfil\cr
$\displaystyle{{\left\LeftPost{par}Q^{N}_{r}\MthAcnt {u}{\matrix{}}
\right\RightPost{par}}{\left\LeftPost{par}\MthAcnt {x}{\matrix{}}
\right\RightPost{par}}={\left\LeftPost{ang}\underline{\underline{Q}}
{\left\LeftPost{par}
\MthAcnt {e}{\matrix{}}_{N}\right\RightPost{par}}
\cdots \underline{\underline{Q}}{\left\LeftPost{par}\MthAcnt {e}{\matrix{}}_{%
1}\right\RightPost{par}}{\ifmmode\cdot\else\.\fi}\MthAcnt {u}{\matrix{}}
{\left\LeftPost{par}\MthAcnt {x}{\matrix{}}+r\MthAcnt {e}{\matrix{}}_{%
1}+\cdots +r\MthAcnt {e}{\matrix{}}_{N}\right\RightPost{par}}
\right\RightPost{ang}}_{\MthAcnt {e}{\matrix{}}_{%
1}\ldots \MthAcnt {e}{\matrix{}}_{N}}
.}$\cr
}}\end{equation}

By considering its spatial Fourier transform
\def\XRefId{}
\begin{equation}\SGMPlab\XRefId\vcenter{\halign{\strut\hfil#\hfil&#\hfil\cr
$\displaystyle{{\left\LeftPost{par}Q^{N}_{r}\MthAcnt {u}{\matrix{}}
\right\RightPost{par}}{\left\LeftPost{par}\MthAcnt {k}{\matrix{}}
\right\RightPost{par}}=\int {\left\LeftPost{par}Q^{%
N}_{r}\MthAcnt {u}{\matrix{}}\right\RightPost{par}}
{\left\LeftPost{par}\MthAcnt {x}{\matrix{}}\right\RightPost{par}}
\exp {\left\LeftPost{par}-i\MthAcnt {k}{\matrix{}}\MthAcnt {%
x}{\matrix{}}\right\RightPost{par}}d^{d}x=\underline{%
\underline{Q}}^{N}_{r}{\left\LeftPost{par}\MthAcnt {%
k}{\matrix{}}\right\RightPost{par}}{\ifmmode\cdot\else\.\fi}\MthAcnt {u}
{\matrix{}}
{\left\LeftPost{par}\MthAcnt {k}{\matrix{}}\right\RightPost{par}}
}$\cr
}}\end{equation}

one can evaluate the limit of vanishing step length
\(
r{\ifmmode\rightarrow\else$\rightarrow$\fi}0\)
 and diverging number of steps
\(
N{\ifmmode\rightarrow\else$\rightarrow$\fi}\infty \)
, while keeping
\(
t=Nr^{2}\)
 fixed. One easily obtains with the central-limit
theorem
\def\XRefId{qtk}
\begin{equation}\SGMPlab\XRefId\vcenter{\halign{\strut\hfil#\hfil&#\hfil\cr
$\displaystyle{\SGMPlim{c}\lim \SGMPdolim _{\stack {N{\ifmmode\rightarrow
\else$\rightarrow$\fi}
\infty }{r{\ifmmode\rightarrow\else$\rightarrow$\fi}0}}\underline{\underline{Q}
}^{N}_{r}{\left\LeftPost{par}\MthAcnt {k}{\matrix{}}
\right\RightPost{par}}}$\hfilneg&$\displaystyle{{}=\underline{\underline{Q}}^{%
t}{\left\LeftPost{par}\MthAcnt {k}{\matrix{}}\right\RightPost{par}}
=\exp {\left\LeftPost{par}-{{t}\over{2d{\left\LeftPost{par}
d+2\right\RightPost{par}}}}{\left\LeftPost{sqb}{\left\LeftPost{par}
2\SGMPmathgrk{a}+d\right\RightPost{par}}\MthAcnt {k}{\matrix{}}^{%
2}\underline{\underline{1}}+2d{\left\LeftPost{par}1-\SGMPmathgrk{a}\right
\RightPost{par}}
\MthAcnt {k}{\matrix{}}\otimes \MthAcnt {k}{\matrix{}}
\right\RightPost{sqb}}\right\RightPost{par}}}$\cr
$\displaystyle{\equiv \exp {\left\LeftPost{par}
\underline{\underline{M}}\right\RightPost{par}}
.}$\cr
}}\end{equation}

After transforming this quantity back into real space, it is easily seen
that the Fourier transform
\def\XRefId{modified_walk}
\begin{equation}\SGMPlab\XRefId\vcenter{\halign{\strut\hfil#\hfil&#\hfil\cr
$\displaystyle{{\left\LeftPost{par}Q^{t}\MthAcnt {u}{\matrix{}}
\right\RightPost{par}}{\left\LeftPost{par}\MthAcnt {x}{\matrix{}}
\right\RightPost{par}}=\int \underline{\underline{Q}}^{%
t}{\left\LeftPost{par}\MthAcnt {r}{\matrix{}}\right\RightPost{par}}
{\ifmmode\cdot\else\.\fi}\MthAcnt {u}{\matrix{}}{\left\LeftPost{par}\MthAcnt
{x}{\matrix{}}
+\MthAcnt {r}{\matrix{}}\right\RightPost{par}}\hskip 0.167em
d^{d}r}$\cr
}}\end{equation}

provides a solution of the following general equation for the vector
field
\(
\MthAcnt {V}{\matrix{}}{\left\LeftPost{par}\MthAcnt {r}{\matrix{}}
,t\right\RightPost{par}}\)

\def\XRefId{time_dependent_problem}
\begin{equation}\SGMPlab\XRefId\vcenter{\halign{\strut\hfil#\hfil&#\hfil\cr
$\displaystyle{{{\SGMPmathgrk{6}\MthAcnt {V}{\matrix{}}}\over{\SGMPmathgrk{6}
t}}={{{\left\LeftPost{par}2\SGMPmathgrk{a}+d\right\RightPost{par}}
\nabla ^{2}\MthAcnt {V}{\matrix{}}+2d{\left\LeftPost{par}
1-\SGMPmathgrk{a}\right\RightPost{par}}\nabla {\left\LeftPost{par}\nabla
{\ifmmode\cdot\else\.\fi}
\MthAcnt {V}{\matrix{}}\right\RightPost{par}}}\over{2
d{\left\LeftPost{par}d+2\right\RightPost{par}}}}}$\cr
}}\end{equation}

The stationary limit
\(
\SGMPmathgrk{6}\MthAcnt {V}{\matrix{}}/\SGMPmathgrk{6}t=0\)
 of this equation obviously provides the Lam\'e
equation if one chooses
\(
\SGMPmathgrk{a}\)
 appropriately
\def\XRefId{}
\begin{equation}\SGMPlab\XRefId\vcenter{\halign{\strut\hfil#\hfil&#\hfil\cr
$\displaystyle{\SGMPmathgrk{a}={{d{\left\LeftPost{par}1-2\SGMPmathgrk{n}
{\left\LeftPost{par}
d-1\right\RightPost{par}}\right\RightPost{par}}}\over{2{\left\LeftPost{par}
d+1-d\SGMPmathgrk{n}{\left\LeftPost{par}d-1\right\RightPost{par}}
\right\RightPost{par}}
}}}$\cr
}}\end{equation}
and identifies
\(
\MthAcnt {V}{\matrix{}}{\left\LeftPost{par}\MthAcnt {r}{\matrix{}}
,t\right\RightPost{par}}\)
with the displacement field
\(
\MthAcnt {u}{\matrix{}}{\left\LeftPost{par}\MthAcnt {r}{\matrix{}}
,t\right\RightPost{par}}\)
.\par

\par In summary, in this formulation the problem of determining a time
dependent Green{'}s function
\(
\underline{\underline{Q}}^{t}{\left\LeftPost{par}\MthAcnt {r}{\matrix{}}
\right\RightPost{par}}\)
, which is used in (\SGMPref{modified_walk}), has
been formulated in terms of a random walk. Obviously eq.
(\SGMPref{elementary_walk}) shows the type of Markov process that one has to
perform in order to solve Lam\'e{'}s equation. Anyhow, we shall show
in the following that although exact, it is not practical to construct
an algorithm directly from (\SGMPref{elementary_walk}).\par

\par A direct and naive implementation of (\SGMPref{elementary_walk})
would involve the following steps. One would launch a random walker at a
boundary site
\(
\MthAcnt {r}{\matrix{}}_{i}=\MthAcnt {x}{\matrix{}}
+\MthAcnt {r}{\matrix{}}\)
 at which the displacement
\(
\MthAcnt {u}{\matrix{}}\)
 is known. The walker would undergo an isotropic
random walk until it touches a crack at point
\(
\MthAcnt {r}{\matrix{}}_{f}=\MthAcnt {x}{\matrix{}}
\)
. During the walk one would calculate the product of
the elementary step matrices
\(
\underline{\underline{Q}}{\left\LeftPost{par}\MthAcnt {e}{\matrix{}}_{%
N}\right\RightPost{par}}\cdots \underline{\underline{Q}}{\left\LeftPost{par}
\MthAcnt {e}{\matrix{}}_{1}\right\RightPost{par}}
\)
 and would finally average at
\(
\MthAcnt {x}{\matrix{}}\)
 over the vectors
\(
\underline{\underline{Q}}{\left\LeftPost{par}\MthAcnt {e}{\matrix{}}_{%
N}\right\RightPost{par}}\cdots \underline{\underline{Q}}{\left\LeftPost{par}
\MthAcnt {e}{\matrix{}}_{1}\right\RightPost{par}}
{\ifmmode\cdot\else\.\fi}\MthAcnt {u}{\matrix{}}{\left\LeftPost{par}
\MthAcnt {r}{\matrix{}}_{%
i}\right\RightPost{par}}\)
 of all incoming random walkers. One would then have
to extrapolate to the limit
\(
r{\ifmmode\rightarrow\else$\rightarrow$\fi}0\)
 {---} which at least in this algorithm necessarily
leads to
\(
N{\ifmmode\rightarrow\else$\rightarrow$\fi}\infty \)
 {---} and according to (\SGMPref{modified_walk})
this procedure should finally yield the required solution.\par

\par But although this process strongly reminds one of the process to
construct the Green{'}s function of the Laplace equation with random
walkers, there are several differences which result in important
difficulties. Unlike in DLA the history of each walk is here extremely
important. The product matrix is explicitly dependent on each individual
step
\(
\MthAcnt {e}{\matrix{}}_{i}\)
 and therefore the path on which the walker travels
from its initial to its final point is extremely important. Thus, the
contribution of each individual walk to the average (\SGMPref{elementary_walk})
is arbitrarily small.\par

\par On the other hand it is necessary to take the limit
\(
N{\ifmmode\rightarrow\else$\rightarrow$\fi}\infty \)
 and thus the number of factors to calculate the
product matrix grows rapidly. But since the eigenvalues of
\(
\underline{\underline{Q}}{\left\LeftPost{par}\MthAcnt {e}{\matrix{}}
\right\RightPost{par}}\)
 are not equal to unity the product matrix
\(
\underline{\underline{Q}}{\left\LeftPost{par}\MthAcnt {e}{\matrix{}}_{%
N}\right\RightPost{par}}\cdots \underline{\underline{Q}}{\left\LeftPost{par}
\MthAcnt {e}{\matrix{}}_{1}\right\RightPost{par}}
\)
 has an uncontrollable norm, which either diverges or
vanishes exponentially fast. From this follows, that also the modulus of
the resulting vector
\(
\underline{\underline{Q}}{\left\LeftPost{par}\MthAcnt {e}{\matrix{}}_{%
N}\right\RightPost{par}}\cdots \underline{\underline{Q}}{\left\LeftPost{par}
\MthAcnt {e}{\matrix{}}_{1}\right\RightPost{par}}
{\ifmmode\cdot\else\.\fi}\MthAcnt {u}{\matrix{}}{\left\LeftPost{par}
\MthAcnt {r_{%
i}}{\matrix{}}\right\RightPost{par}}\)
 is uncontrollable. Similar arguments show immediately
that not only the modulus, but also the direction of the resulting
vector is uncontrollable. The consequence of these considerations is,
that a naive implementation of the algorithm suggested by
(\SGMPref{elementary_walk}) results in enormous statistical fluctuations of
both the modulus and the direction of
\(
{\left\LeftPost{par}Q^{N}_{r}\MthAcnt {u}{\matrix{}}
\right\RightPost{par}}{\left\LeftPost{par}\MthAcnt {x}{\matrix{}}
\right\RightPost{par}}\)
, which cannot be suppressed in numerical
simulations.\par

\par Another important point is, that the previously described algorithm
neglects the presence of a crack, which substantially changes the stress
field. It is also not easily possible to include the boundary condition
at the crack (\SGMPref{boundary}), because it is expressed in terms of
the stress field, which is not included in the previous derivation. Of
course one could reformulate the boundary condition in terms of the
displacement field, but this requires spatial derivatives of the
displacement field which are even less accessible than the displacement
itself.\par

\par However, the previous considerations showed that one purpose of the
random walkers is to construct the matrix
\(
\underline{\underline{Q}}^{t}{\left\LeftPost{par}\MthAcnt {r}{\matrix{}}
\right\RightPost{par}}\)
, which is the basis of (\SGMPref{modified_walk}).
But in an infinite system this matrix, which is actually already the
average of all product matrices of walks starting at the origin and
terminating at
\(
\MthAcnt {r}{\matrix{}}\)
, can be calculated analytically. Thus, one could
evaluate (\SGMPref{modified_walk}) directly using random walks, which
would then only be used to determine the initial and final positions
\(
\MthAcnt {x}{\matrix{}}\)
 and
\(
\MthAcnt {x}{\matrix{}}+\MthAcnt {r}{\matrix{}}
\)
. Here, the history of each individual walk is no
longer relevant and thus one does not obtain the huge fluctuations
described before.\par

\par
\(
\underline{\underline{Q}}^{t}{\left\LeftPost{par}\MthAcnt {r}{\matrix{}}
\right\RightPost{par}}\)
 is obtained by a Fourier transformation of (\SGMPref{qtk}) into real space
(see Appendix). Finally one finds
\def\XRefId{}
\begin{equation}\SGMPlab\XRefId\vcenter{\halign{\strut\hfil#\hfil&#\hfil\cr
$\displaystyle{\underline{\underline{Q}}^{t}{\left\LeftPost{par}\MthAcnt {r
}{\matrix{}}\right\RightPost{par}}={{1}\over{%
2\SGMPmathgrk{p}}}{\left\LeftPost{sqb}f_{s}{\left\LeftPost{par}r
,t\right\RightPost{par}}+f_{a}{\left\LeftPost{par}r,t\right\RightPost{par}}
\right\RightPost{sqb}}{\ifmmode\cdot\else\.\fi}\underline{\underline{1}}-{{1}
\over{2\SGMPmathgrk{p}}}f_{a}{\left\LeftPost{par}r,t\right\RightPost{par}}
{\ifmmode\cdot\else\.\fi}2{{\MthAcnt {r}{\matrix{}}\otimes \MthAcnt
{r}{\matrix{}}
}\over{\MthAcnt {r}{\matrix{}}^{2}}}
}$\cr
}}\end{equation}

where
\def\XRefId{}
\begin{equation}\SGMPlab\XRefId\vcenter{\halign{\strut\hfil#\hfil&#\hfil\cr
$\displaystyle{f_{s}{\left\LeftPost{par}r,t\right\RightPost{par}}={{%
\exp {\left\LeftPost{par}-r^{2}/4\SGMPmathgrk{l}_{1}
t\right\RightPost{par}}}\over{4\SGMPmathgrk{l}_{1}t}}
+{{\exp {\left\LeftPost{par}-r^{2}/4\SGMPmathgrk{l}_{%
2}t\right\RightPost{par}}}\over{4\SGMPmathgrk{l}_{2}t}}
}$\cr
$\displaystyle{f_{a}{\left\LeftPost{par}r,t\right\RightPost{par}}=
{\left\LeftPost{par}
1+{{4\SGMPmathgrk{l}_{1}t}\over{r^{2}}}
\right\RightPost{par}}{{\exp {\left\LeftPost{par}-r^{%
2}/4\SGMPmathgrk{l}_{1}t\right\RightPost{par}}}\over{4
\SGMPmathgrk{l}_{1}t}}-{\left\LeftPost{par}1+{{4\SGMPmathgrk{l}_{%
2}t}\over{r^{2}}}\right\RightPost{par}}
{{\exp {\left\LeftPost{par}-r^{2}/4\SGMPmathgrk{l}_{%
2}t\right\RightPost{par}}}\over{4\SGMPmathgrk{l}_{2}t}}
}$\cr
}}\end{equation}

This result has to be commented in several respects.\par

\par It is clearly seen from the form of the two functions
\(
f_{s}\)
 and
\(
f_{a}\)
, that the underlying elementary process is a random
walk. They consist of two Gaussians, which are the result of the
diffusion of free particles launched at the origin of an infinite
domain. The limiting case of the pure Laplacian is obtained by setting
\(
\SGMPmathgrk{n}{\ifmmode\rightarrow\else$\rightarrow$\fi}-\infty \)
. In this case
\(
\underline{\underline{Q}}^{t}{\left\LeftPost{par}\MthAcnt {r}{\matrix{}}
\right\RightPost{par}}\)
 simplifies to the Green{'}s function of the
corresponding Laplace problem, which is a pure Gaussian.\par

\par Since one requires
\(
t=Nr^{2}\)
 to be fixed, one has introduced a new, free time
variable. But none of the recent and efficient implementations of DLA
algorithms uses an explicit time definition. In DLA the time it takes
for a walker to reach a cluster is not relevant. Here, \(k\)
explicitly enters
\(
\underline{\underline{Q}}^{t}{\left\LeftPost{par}\MthAcnt {r}{\matrix{}}
\right\RightPost{par}}\)
.\par

\par The previously calculated
\(
\underline{\underline{Q}}^{t}{\left\LeftPost{par}\MthAcnt {r}{\matrix{}}
\right\RightPost{par}}\)
 is only valid in an infinite domain. In (\SGMPref{elementary_walk}) one
averages essentially over all possible random
walks that connect the origin with the point
\(
\MthAcnt {r}{\matrix{}}\)
. But in the presence of a crack, many of the walks
terminate at the crack before reaching
\(
\MthAcnt {r}{\matrix{}}\)
. Also those walks are taken into account in eqs.
(\SGMPref{elementary_walk}-\SGMPref{modified_walk}) to calculate
\(
\underline{\underline{Q}}^{t}{\left\LeftPost{par}\MthAcnt {r}{\matrix{}}
\right\RightPost{par}}\)
. This deficiency could be altered by considering only
those walks, which do not leave a restricted area, like in calculations
of {``}first-passage-times{''}, but this case cannot be calculated
analytically.\par

\par Equation (\SGMPref{modified_walk}) leads to the solution of the
Lam\'e equation only in the stationary limit
\(
t{\ifmmode\rightarrow\else$\rightarrow$\fi}\infty \)
. But in this limit
\(
\underline{\underline{Q}}^{t}{\left\LeftPost{par}\MthAcnt {r}{\matrix{}}
\right\RightPost{par}}\)
 vanishes identically, which means that one only
obtains the trivial solution
\(
\MthAcnt {u}{\matrix{}}{\left\LeftPost{par}\MthAcnt {r}{\matrix{}}
\right\RightPost{par}}\equiv 0\)
.\par

\par As a result one obtains, that also a numerical simulation of
(\SGMPref{modified_walk}) {---} although technically possible {---}
cannot be done, since mainly the time variable \(t\) cannot be
interpreted in a simulation.\par

\par The essential result of the previous considerations is, that the
matrix
\(
\underline{\underline{Q}}^{t}{\left\LeftPost{par}\MthAcnt {r}{\matrix{}}
\right\RightPost{par}}\)
 plays the role of a Green{'}s function of the time
dependent problem (\SGMPref{time_dependent_problem}) and the original
algorithm simply provides a prescription how to construct it using
random walks. On the other hand, one would in all cases end up with a
Green{'}s function in an infinite domain, which in fact can be calculated
directly and analytically.\par

\def\XRefId{Algorithm}\section{\SGMPlab\XRefId Algorithm}

\par From the previous sections it is clear that the use of the original
random walker method always ends up in the calculation of a Green{'}s
function for the displacement field. Both, the boundary condition at the
crack surface and the breaking criterion, are usually given in terms of
the stress field which is related to the spatial derivative of the
displacement field using Hooke{'}s law. Thus it seems natural to calculate
the Green{'}s function for the stress field analytically and use this to
construct a random walker algorithm.\par

\par Therefore, consider an infinite domain in the absence of cracks
which is kept fixed at infinity. At the origin one applies a point force
\(
\MthAcnt {F}{\matrix{}}\)
 which generates a stress field
\(
\underline{\underline{\SGMPmathgrk{s}}}^{*}{\left\LeftPost{par}\MthAcnt {%
r}{\matrix{}}\right\RightPost{par}}\)
 throughout the whole system and is governed by
\(
\hbox{\rm Div}\hskip 0.212em \underline{\underline{\SGMPmathgrk{s}}}^{%
*}=\MthAcnt {F}{\matrix{}}{\ifmmode\cdot\else\.\fi}
\SGMPmathgrk{d}{\left\LeftPost{par}
\MthAcnt {r}{\matrix{}}\right\RightPost{par}}\)
. Using the method of Kolossov-Mushkelishvili
\SGMPcite{fung,kolossov_mushkelishvili}{} we can
calculate the stress field in the whole system and obtain
\def\XRefId{greens_function_org}
\begin{equation}\SGMPlab\XRefId\vcenter{\halign{\strut\hfil#\hfil&#\hfil\cr
$\displaystyle{\SGMPmathgrk{s}^{*}_{xx}=2\Re {\left\LeftPost{par}{{%
c}\over{z}}\right\RightPost{par}}+\Re {\left\LeftPost{par}
{{\overline{z}c}\over{z^{2}}}+\SGMPmathgrk{c}
{{\overline{c}}\over{z}}\right\RightPost{par}}
}$\cr
$\displaystyle{\SGMPmathgrk{s}^{*}_{yy}=2\Re {\left\LeftPost{par}{{%
c}\over{z}}\right\RightPost{par}}-\Re {\left\LeftPost{par}
{{\overline{z}c}\over{z^{2}}}+\SGMPmathgrk{c}
{{\overline{c}}\over{z}}\right\RightPost{par}}
}$\cr
$\displaystyle{\SGMPmathgrk{s}^{*}_{xy}=-\Im {\left\LeftPost{par}{{%
\overline{z}c}\over{z^{2}}}+\SGMPmathgrk{c}{{%
\overline{c}}\over{z}}\right\RightPost{par}}}$\cr
}}\end{equation}

 where we represent all 2d vectors as complex numbers
\def\XRefId{}
\begin{equation}\SGMPlab\XRefId\vcenter{\halign{\strut\hfil#\hfil&#\hfil\cr
$\displaystyle{z=x+iy}$\cr
$\displaystyle{c={{F_{x}+iF_{y}}\over{2\SGMPmathgrk{p}{\left\LeftPost{par}
1+\SGMPmathgrk{c}\right\RightPost{par}}}}}$\cr
$\displaystyle{\SGMPmathgrk{c}=3-4\SGMPmathgrk{n}}$\cr
}}\end{equation}

 and
\(
\Re {\left\LeftPost{par}z\right\RightPost{par}}\)
 (
\(
\Im {\left\LeftPost{par}z\right\RightPost{par}}\)
) denote the real (imaginary) part of
\(
z\)
. Now one has to take into account that the
probability for a random walker, that is launched from the center of a
circle of radius
\(
r\)
, to reach a certain element
\(
d\SGMPmathgrk{4}\)
 at the perimeter of this circle is
\(
p\hskip 0.167em d\SGMPmathgrk{4}=d\SGMPmathgrk{4}/2\SGMPmathgrk{p}r\)
. This term has to be divided out of (\SGMPref{greens_function_org}) and
therefore we define
\def\XRefId{greens_function}
\begin{equation}\SGMPlab\XRefId\vcenter{\halign{\strut\hfil#\hfil&#\hfil\cr
$\displaystyle{\underline{\underline{\SGMPmathgrk{s}}}=
2\SGMPmathgrk{p}{\left\LeftPost{vb}z\right\RightPost{vb}}
{\ifmmode\cdot\else\.\fi}\underline{\underline{\SGMPmathgrk{s}}}^{*}.}$\cr
}}\end{equation}

\par

\par Using this Green{'}s function
\(\underline{\underline \sigma}\), which is the basis of the
following algorithm, we define a new method to determine the stress
field in the medium:

\SGMPbeginList{numeral}{}

\SGMPitem\def\XRefId{}\SGMPlab\XRefId Since (\SGMPref{equilibrium}) is linear,
we can obtain the
solution for a line of forces as a superposition of elementary point
forces. Like in DLA, where each point which is kept at a nonzero
potential is interpreted as a source of walkers, each point force acts
also here as a source of random walkers. Random walkers are launched
{---} like in DLA one at a time {---} with equal probabilities from
each point
\(
\MthAcnt {r}{\matrix{}}_{i}\)
 where a force
\(
\MthAcnt {F}{\matrix{}}_{i}\)
 is applied. All forces are usually of unit strength.
Each walker undergoes an ordinary random walk until it touches a
particle of the cluster located at point
\(
\MthAcnt {r}{\matrix{}}_{f}\)
. Afterwards one calculates the stress tensor
\(
\underline{\underline{\SGMPmathgrk{s}}}{\left\LeftPost{par}\MthAcnt
{r}{\matrix{}}_{%
f}-\MthAcnt {r}{\matrix{}}_{i},\MthAcnt {F}{\matrix{}}_{%
i}\right\RightPost{par}}\)
 according to (\SGMPref{greens_function}) and
accumulates this stress in the counter of the appropriate surface
element (see below).

\SGMPitem\def\XRefId{}\SGMPlab\XRefId A crack is represented as an aggregate of
particles of unit
diameter. Like in DLA the growth of a crack has to be initiated by
locating a seed particle at some point, which then acts as a single
microcrack, from which the crack grows. Since we expect strong
fluctuations during the growth of the crack we use a special scheme to
reduce the fluctuations and get a better estimate of the stress field.
It is well known from DLA that the introduction of anisotropic
noise-reduction schemes
\SGMPcite{eckmann_meakin_procaccia_zeitak,derrida_hakim_vannimenus}{},
immediately leads to strong lattice effects. Although similar schemes
have not shown such effects in lattice models of crack growth
\SGMPcite{fernandez_guinea_louis}{} we try to avoid possible
problems, and therefore our scheme must not define any surface nor
prefer any direction. The crack is represented by the particles as
indicated in figure 1: Each
particle represents two stress counters, one for each crack surface.
Only the particles at the tip of a branch, i. e. those with only one
parent particle but no children, have one single counter. In these
counters the stresses \(\underline{\underline \sigma}\) carried
by the incoming walkers are accumulated.

\SGMPitem\def\XRefId{}\SGMPlab\XRefId Obviously the correct boundary condition
(\SGMPref{boundary}) is
not obeyed by simply calculating stresses according to
(\SGMPref{greens_function}), but an unbalanced force
\[\vec F_{\hbox{unbal}}=\underline{\underline{\sigma}}\cdot \vec n \neq 0\]%
remains. Therefore, we apply a virtual balancing force
\def\XRefId{}
\begin{equation}\SGMPlab\XRefId\vcenter{\halign{\strut\hfil#\hfil&#\hfil\cr
$\displaystyle{\MthAcnt {F}{\matrix{}}_{\hbox{\rm bal}}=-
\underline{\underline{\SGMPmathgrk{s}}}{\ifmmode\cdot\else\.\fi}
\MthAcnt {n}{\matrix{}}
}$\cr
}}\end{equation}

to the appropriate surface element at
\(
\MthAcnt {r}{\matrix{}}_{f}\)
, which exactly compensates the spurious shear
component
\(
\SGMPmathgrk{t}\)
 and stress component perpendicular to the crack
surface
\(
\SGMPmathgrk{s}_{\perp }\)
. Thus it is in principle sufficient to accumulate at
the crack only the stress component parallel to the crack surface
\(
\SGMPmathgrk{s}_{\parallel }\)
. But on the other hand we will show later that the
shear component  will be essential to determine the direction into which
the crack grows and thus we accumulate at
\(
\MthAcnt {r}{\matrix{}}_{f}\)
 the shear component as well as
\(
\SGMPmathgrk{s}_{\parallel }\)
. As was said above, a random walker is started at
each point where a force is applied. Consequently, the walker has to be
restarted from
\(
\MthAcnt {r}{\matrix{}}_{f}\)
 and is reinserted into the system, this time with the
new force
\(
\MthAcnt {F}{\matrix{}}_{\hbox{\rm bal}}\)
. This process of constantly reinserting the walker
into the system is repeated until the modulus of
\(
\MthAcnt {F}{\matrix{}}_{\hbox{\rm bal}}\)
 drops below a certain threshold, which in our
simulations is typically chosen as \(10^{-8}\). Afterwards the
walker is discarded and a new walker is started. This repeated process
can be interpreted as a method to relax the stress at the surface and
can best be understood in terms of a boundary integral method
\SGMPcite{kessler_koplik_levine}{}. Physically this process is
like a multiple scattering method. A similar situation occurs in DLA in
simulations of viscous fingering with nonvanishing surface tension
\SGMPcite{kadanoff,vicsek_prl_1984,meakin_family_vicsek}{} or in
simulations of electrochemical deposition with nonvanishing surface
impedance \SGMPcite{halsey_leibig}{}. Both cases are
mapped to a simulation of sticking probability DLA, in which the walker
stops at the cluster with a probability smaller than unity.

\SGMPitem\def\XRefId{}\SGMPlab\XRefId Next we have to define a growth rule.
Here it has to be stressed
that the growth of the crack must not be governed by the fact that the
crack is touched by a walker, but only the accumulated stress tensor
should determine the growth of the crack. Since the stress tensor is
symmetric, it defines eigenvalues and eigenvectors, which determine the
principal stresses and principal stress directions. The eigenvalues also
allow to distinguish between tensile and compressive stress, because
both have a different sign. Now, we provide two material parameters: one
for tension
\(
\SGMPmathgrk{s}_{t}\)
 and one for compression
\(
\SGMPmathgrk{s}_{c}\)
. They represent the material strength under the
corresponding load. Especially the strength under tensile load
\(
\SGMPmathgrk{s}_{t}\)
 can be related to the intermolecular cohesion forces.
If either of them is exceeded by one eigenvalue, the crack grows and a
new particle is added to the crack. The direction into which to put the
new particle is determined by the eigenvector of the other eigenvalue.
This breaking rule is chosen such that for a pure uniaxial tension the
crack grows perpendicular to the direction of the force. In this growth
rule only the principal stresses, which are purely tensile or
compressive, are checked against the material strength. This situation
is similar to the growth rule in the central-force model \SGMPcite{feng_sen}{},
in which also no shear or bending modes
are used to break a bond.

\SGMPitem\def\XRefId{}\SGMPlab\XRefId Once the crack has grown it has to
release its stress. Here, we
study two different situations. In the first one the stress is only
released locally, i. e. only the stress counter of the particle to which
a new particle is added is cleared, while all other counters keep their
values. This situation can physically be compared to a material in which
fatigue plays an important role since even a small load can accumulate
and break a sample if it is acting long enough. In the second situation
we study a global relaxation. In this scheme we not only clear the
counter of the particle at which the crack has grown, we also clear all
counters in the entire cluster after the crack has grown by
\(M\) particles. The parameter \(M\) is a fatigue
parameter. The first situation can be expressed in terms of this
parameter as
\(
M=\infty \)
{}.

\SGMPendList

\def\XRefId{Results}\section{\SGMPlab\XRefId Results}

\par
In the following we are going to present the results obtained by
simulating the previous method. We simulated several different boundary
conditions: uniaxial tension and compression, isotropic tension and pure
shear (fig. ). In order to compare with
experimental results we also simulated a four point shear geometry
 and another
geometry used in experiments . We furthermore studied
the influence of material strength and fatigue.

\par The length scale in all following simulations is the particle
diameter and all applied forces have unit length, which defines the unit
of the stress.\par

\def\XRefId{}\subsection{\SGMPlab\XRefId Basic results}

\par In the following we will present the results of three different
simulations with the boundary conditions uniaxial tension, isotropic
tension and pure shear. The material strengths are typically chosen to
be
\(
\SGMPmathgrk{s}_{t}=20\)
 and
\(
\SGMPmathgrk{s}_{c}=-20\)
 (sc. the applied forces are of unit size). Thus, the
system breaks equally well under compression and under tension. Since
the stress components parallel to the crack surface of one walker are
typically around
\(
0.5\)
, at least
\(
N=40\)
 walker have to touch a certain site before the
cluster can grow. As the seed of all clusters we place two particles
with unit diameter at the points (0,0.5) and (0,-0.5). This initial
geometry is chosen arbitrarily, but simulations with other geometries
show that the results  depend neither on the number nor on the specific
geometry of these seed particles. The Poisson ratio is
\(\SGMPmathgrk{n}=0.2\)
 and the fatigue parameter is \(M=10\). All
clusters have a size of \(6,000\) particles.\par

In figure 3 we show a typical result of a
simulation of uniaxial tension. Here, forces are located along the lines
\(\vec r=(\pm 2000,y),\quad y\in [-2000,2000] \)
 and point into the positive (negative)
\(x\)
 direction. One obtains essentially a one-dimensional,
straight crack without side branches that grows perpendicular to the
direction of the force. This result is physical and can easily be
observed in experimental and other numerical studies.

In figure 4 we show a typical result of a
simulation of pure shear. In this simulation all forces are again
located along the lines
\(r=(\pm 2000,y)\).
But now the forces are pointing into the positive (negative)
\(y\)
direction to produce a pure shear field. Although we
use the same microcrack as in the previous simulation the shape of the
crack is now drastically altered. The resulting cracks show a pronounced
cross like shape. This main shape is a result of the pure shear field,
whose stress tensor contains only off-diagonal elements, and thus the
eigenvectors always point into a $45^\circ$ angle from a surface. We have
to remind, that this shape is not a lattice effect but a consequence of
the applied forces.

\par The reason for the symmetry of the cross-like shape is that the
material strengths for tension and compression are symmetric,
\(\SGMPmathgrk{s}_{c}=-\SGMPmathgrk{s}_{t}\). The case of asymmetric strength
will be shown later.
The cross-like shape under symmetric breaking conditions has been
observed in other numerical studies
\SGMPcite{meakin_li_sander_louis_guinea,hinrichsen_hansen_roux}{}
on much smaller length scales. In experiments this shape is usually not
observed, because the material strength for tension is usually smaller
than for compression. \par

\par One also observes small side branches which form right angles with
the main branch. We have to stress, that these right angles cannot be
the result of lattice effects because the whole method is formulated
off-lattice. Our simulations are the first off-lattice simulations which
show this behavior. The formation of right angles is completely
determined by the cracking process and has a physical origin: Since
there are no restoring forces acting on the crack surface, the boundary
condition (\SGMPref{boundary}) only allows stresses parallel to the
surface. Consequently side branches can only grow perpendicular to the
main branches in the vicinity of the main crack.\par

A typical result for a crack pattern under isotropic tension is shown in
figure 5. Here, the radially outward pointing
forces are located along a circle of radius
\(2000\). One obtains a ramified structure which can clearly
be distinguished from DLA: The typical tip-splitting instability
vanished and the side branches show again a tendency to grow
perpendicular to their main branches.

\def\XRefId{}\subsection{\SGMPlab\XRefId Fatigue}

\par As a first variation we want to study the influence of the fatigue
parameter by setting
\(M=\infty \)
 while keeping all other parameters fixed. Physically
we simulate now a material in which a small load can accumulate and can
damage the material. Eventually this can lead to the formation of a
large crack.\par

\par Since we use
\(M=\infty \)
 we need much less walkers to reach the material
strength and can simulate much larger cracks. Here, we show cracks
containing \(20,000\) particles.\par

Under uniaxial tension (fig. 6) we obtain again
a narrow straight crack perpendicular to the applied load, but we
observe the formation of many small side branches, which grow
perpendicular to the main crack and which can themselves form
perpendicular side branches. We have to stress again that the formation
of perpendicular side branches is not a result of lattice effects but
quite physical. However, the formation of multiple side branches is not
physical and thus we obtain, that the fatigue parameter is essential to
describe the physical processes.

Under shear load (fig. 7) we observe the formation
of the cross-like main branches. Like in the previous case the
introduction of fatigue leads also in this case to the formation of many
small side branches which again may form side branches themselves. This
leads eventually to a dendritic shape of the crack.

Very interesting is the result under isotropic tension (fig. 8). Like in the
previous cases we obtain a ramified
main crack which is covered with many small side branches. Again we
observe that theses side branches grow perpendicular to the main branch.
This fact is particularly interesting in this geometry since here no
growth direction is favored by the boundary condition and thus the right
angles can only be explained by the previously given physical arguments.
Furthermore one has to notice the stability of the crack tips which do
not show a tendency to split. One rather observes the appearance of side
branches behind the crack tip.

\def\XRefId{}\subsection{\SGMPlab\XRefId Variation of the material strength}

\par In this section  we want to discuss the effect of the material
strength on the results of the simulation. We set the strength to two
extreme values: either to zero or to
\(
{\ifmmode\pm\else$\pm$\fi}100\)
 while keeping
\(
M=\infty \)
 fixed.\par

\par In the figures
9 {--} 11 the
results for
\(
\SGMPmathgrk{s}_{c}=\SGMPmathgrk{s}_{t}=0\)
 are shown. Setting both values to zero means, that
the crack grows as soon as it is touched by a walker.\par

In the uniaxial tension experiment (fig. 9) the crack remains its elongated
shape but
the number and shape of the side branches changed drastically. They
completely lost their straight appearance and are much more ramified.
Also the pronounced $90^\circ$ angles vanish. Now, the crack looks very
much like a DLA cluster with anisotropic sticking probability .

In the shear simulation (fig. 10) the
global cross shape of the crack is still present. But here  in
contrast to the uniaxial case  the right angles between main and
side branches and the very straight shape of all branches are still
present. Amazingly, the dendritic shape of the cracks under shear is
extremely stable and is not destroyed even in the limit of zero
strength.

In the case of isotropic tension (fig. 11) almost all features emphasized above
are
gone. The cluster does no longer reveal side branching and $90^\circ$
angles between branches or screening of curved cracks. The resulting
crack looks like a typical DLA cluster of the corresponding size.

\par As a result one obtains, that in the limit of zero strength our
model crosses over into a anisotropic DLA like behavior. This is readily
explained because in the limit of zero strength the noise generated by
the random walkers dominates over the structure imposed by the outer
boundaries. But the vectorial character of the problem is not completely
gone. It introduces an intrinsic anisotropy as is clearly seen by
comparing the uniaxial and the shear case: In the first case one still
obtains one straight crack perpendicular to the force, whereas in the
latter case one obtains the cross like shear crack. Such a peculiarity
cannot be found in the scalar DLA.\par

In the figures 12--14  the results for
\(\sigma_c=\sigma_t=100\)
 are shown. In the shear simulation (fig. 12) one observes that the dendritic
shape of the
crack is extremely stable. Even in the limit of high noise reduction the
dendritic shape of the four main branches does not change. Neither the
number nor the size of the main branches seems to decrease. Also in the
case of uniaxial tension (fig. 13) the
overall shape of the cluster does not change. One still observes a
straight structure with many side branches and even the side branches
that grow parallel to the main branch remain. In the isotropic geometry
(fig. 14) one observes that the main
and the side branches tend to be straighter and the overall crack seems
to be less ramified. Here one can see clearly the tendency of side
branches to grow perpendicular to the main branches.

\def\XRefId{}\subsection{\SGMPlab\XRefId Asymmetric strength}

\par Here we want to address the question of asymmetric strengths. In
typical real material like concrete one usually finds the behavior
\(
{\left\LeftPost{vb}\SGMPmathgrk{s}_{c}\right\RightPost{vb}}
{\ifmmode>\else$>$\fi}{\left\LeftPost{vb}
\SGMPmathgrk{s}_{t}\right\RightPost{vb}}\)
. Thus, as special cases we want to comment on the
situation, that the material only breaks under tension:
\(
\SGMPmathgrk{s}_{c}=-\infty ,\hskip 0.212em \SGMPmathgrk{s}_{t}
=20\)
. We also studied the opposite case in which the
medium breaks only under compression:
\(
\SGMPmathgrk{s}_{t}=\infty ,\hskip 0.212em \SGMPmathgrk{s}_{c}
=-20\)
.\par

The most obvious change in the result is obtained in the shear
experiments (fig. 15). As soon as one
strength is chosen much larger than the other one, only one of the
diagonal main branches remains. The curvature of the crack is purely a
boundary effect. In fact, according to the breaking mode either one or
the other diagonal is selected. This is in complete agreement with
previous simulations . One also finds that
the pronounced dendritic shape that was observed earlier has completely
vanished and the single crack does no longer contain any side branch at
all.

A similar observation can be made in the uniaxial tension simulation
(fig. 16). If one only allows
breaking under tension one obtains a single straight crack perpendicular
to the force direction. Again, all side branches disappear. In the other
case, in which only the breaking under compression is allowed, the crack
perpendicular to the load direction disappears and a single crack
parallel to the load direction is observed. This effect is readily
understood with the Young experiment: A tensile load on an elongated bar
will not only increase the length of the bar but it will also influence
the width of the bar, according to the Poisson ratio. Thus, the breaking
mode observed here is a result of the positive Poisson ratio.

The suppression of compressive breaking seems to have the smallest
effect in the isotropic case (fig. 17). Here, there seems to be no
difference between a crack generated only for tensile breaking and a
crack generated under tensile and compressive cracking: One observes
$90^\circ$ angles between main and side branches, one observes side
branching rather than tip splitting and one obtains the screening
behavior described earlier.

\def\XRefId{}\subsection{\SGMPlab\XRefId Experiments}

\par We also reproduce with our new model real experiments performed on
concrete. As an example we want to show the results obtained for the
{``}four-point-shear experiment{''} \SGMPcite{vanmier_schlangen_noorumohamed}{}
and a mixed
{``}tensile-shear experiment{''} \SGMPcite{vanmier_noorumohamed_timmers}{}.\par

In the four point shear experiment a concrete sample  containing
two notches in it to initiate the fracture process at well defined
points  is loaded with four shear forces as indicated in figure 18. One has to
mention, that in the
experiments of course finite samples are used, which imposes additional
boundaries on the simulation. In our simulation we simplify this and
apply four forces to an infinite plane. But we showed with a
boundary-integral calculation of the stress field that this
simplification does not lead to substantial difficulties. For
simplification we describe the notches by two single particles at the
respective positions (fig. 18). The
geometrical relations for the location and the magnitude of the forces
are the same as in the experiment.

The result of our simulation is shown in figure 19, whereas the result of
experiments is
shown in figure 20. Both results
show the same behavior: One observes curved cracks, which start with an
almost $45^\circ$ angle and cross over into a straight behavior: In the
initial stages of growth the crack develops according to a pure shear
field generated by the two forces between which the microcracks are
located. This results in an roughly $45^\circ$ angle. If the cracks get
larger the situation changes and growth occurs according to a
compressive field generated by the two forces located near the
notch.

\par In the experiment one obtains in the end one large and one small
crack, while we observe in the simulations a symmetric situation. One
observes experimentally that initially two cracks start to grow, but one
of them stops after one has reached the yield stress
\SGMPcite{vanmier_schlangen_noorumohamed}{}. This yield stress is
not accessible in our simulations, which is the reason for the symmetric
crack pattern.\par

In the second calculation we simulated another experiment also performed
on a concrete sample which is known as load-path 2: To a
concrete sample first a tensile load is applied until a crack of a
certain size developed, then it is unloaded and a tensile shear load is
applied. The result of our simulation (fig. 21) of this experiment is compared
to the
experimental results (fig. 22). One
observes first a long straight tensile crack as a result of the tensile
load. During the second load phase the diagonal shear cracks
emerge.

\def\XRefId{}\subsection{\SGMPlab\XRefId Biased walks}

\par In all simulations described above the walkers performed an
isotropic random walk. This is in agreement with the calculations of
Roux. In this section we want to consider the case of biased random
walks, in which each walker jumps into a preferred direction.\par

\par One physical motivation of the bias is the following. The stress
\(\underline{\underline\sigma}\) and the strain field
\(\underline{\underline\varepsilon}\) lead inside an elastic
medium to an elastic energy \(U\) which is stored in the fields
\def\XRefId{}
\begin{equation}\SGMPlab\XRefId\vcenter{\halign{\strut\hfil#\hfil&#\hfil\cr
$\displaystyle{U={{1}\over{2}}\SGMPmathgrk{3}_{ij}{\ifmmode\cdot\else\.\fi}
\SGMPmathgrk{s}_{ij}}$\cr
}}\end{equation}

In our case of an infinite and isotropic elastic medium with a point
force \(\vec F\) at the origin this can be calculated at point
\(\vec r\) as
\def\XRefId{potential_energy}
\begin{equation}\SGMPlab\XRefId\vcenter{\halign{\strut\hfil#\hfil&#\hfil\cr
$\displaystyle{U{\left\LeftPost{par}\MthAcnt {r}{\matrix{}}
\right\RightPost{par}}
\propto 1+2{{{\left\LeftPost{par}2-\SGMPmathgrk{n}\right\RightPost{par}}
{\left\LeftPost{par}1+\SGMPmathgrk{n}\right\RightPost{par}}}
\over{{\left\LeftPost{par}
1-\SGMPmathgrk{n}\right\RightPost{par}}^{2}}}{\ifmmode
\cdot\else\.\fi}\SGMPlim{r}\cos \SGMPdolim
^{2}\SGMPmathgrk{4}}$\cr
}}\end{equation}

where \(\cos \varphi \propto \vec F\cdot\vec r\). Now, we
consider the regions with high elastic energy as hot regions in which
crack growth is most likely. Thus it is reasonable to let the walker
preferentially diffuse into these regions by using an angular jump
probability of the type
\def\XRefId{}
\begin{equation}\SGMPlab\XRefId\vcenter{\halign{\strut\hfil#\hfil&#\hfil\cr
$\displaystyle{p{\left\LeftPost{par}\SGMPmathgrk{4}
\right\RightPost{par}}\propto 1+\SGMPmathgrk{a}
{\ifmmode\cdot\else\.\fi}\SGMPlim{r}\cos \SGMPdolim ^{2}\SGMPmathgrk{4}
}$\cr
}}\end{equation}

where \(\alpha\) is now a free model parameter and
\(\varphi\) is the angle between jump and force
direction.\par

\par Since we must not overestimate the contribution of each walker to
the stress counters at the crack surface, we now have to accumulate
scaled stresses
\def\XRefId{}
\begin{equation}\SGMPlab\XRefId\vcenter{\halign{\strut\hfil#\hfil&#\hfil\cr
$\displaystyle{\MthAcnt {\underline{\underline{\SGMPmathgrk{s}}}}{\tilde }
{\left\LeftPost{par}\MthAcnt {r}{\matrix{}}_{i}-\MthAcnt {%
r}{\matrix{}}_{f},\MthAcnt {F}{\matrix{}}
\right\RightPost{par}}={{\underline{\underline{\SGMPmathgrk{s}}}
{\left\LeftPost{par}
\MthAcnt {r}{\matrix{}}_{i}-\MthAcnt {r}{\matrix{}}_{%
f},\MthAcnt {F}{\matrix{}}\right\RightPost{par}}}
\over{p{\left\LeftPost{par}\SGMPmathgrk{4}\right\RightPost{par}}}}}$\cr
}}\end{equation}

where
\(
\cos \SGMPmathgrk{4}\propto {\left\LeftPost{par}\MthAcnt {r}{\matrix{}}_{%
i}-\MthAcnt {r}{\matrix{}}_{f}\right\RightPost{par}}
{\ifmmode\cdot\else\.\fi}\MthAcnt {F}{\matrix{}}\)
. Here we take into account the anisotropic
distribution of walkers.\par

\par In this formulation the walkers diffuse preferentially into the
direction of the force, which is certainly a good choice for walkers
that are launched at the outer boundary. But if one considers the
walkers that are relaunched at the crack surface, this might lead to
unphysical effects: Since all walkers diffuse until they touch the
crack, the relaunched walkers might touch the crack again in an angle
\(
\SGMPmathgrk{4}\approx \SGMPmathgrk{p}/2\)
 which finally leads to
\(
\MthAcnt {\underline{\underline{\SGMPmathgrk{s}}}}{\tilde }
{\ifmmode\rightarrow\else$\rightarrow$\fi}
\infty \)
. Thus, we still overestimate the contribution of
individual walkers. This shows, that the bias does not only increase the
density of walkers in certain regimes of space, but it also shifts the
balance between the walkers launched at the outer boundary and at the
crack. Therefore, we also consider a second possibility of a bias, in
which the walkers preferentially diffuse perpendicular to the force
direction
\def\XRefId{}
\begin{equation}\SGMPlab\XRefId\vcenter{\halign{\strut\hfil#\hfil&#\hfil\cr
$\displaystyle{p{\left\LeftPost{par}\SGMPmathgrk{4}
\right\RightPost{par}}\propto 1+\SGMPmathgrk{a}
{\ifmmode\cdot\else\.\fi}\SGMPlim{r}\sin \SGMPdolim ^{2}\SGMPmathgrk{4}
}$\cr
}}\end{equation}

\par

The figures 23  and 24  show
typical crack structures generated with a cosine-type and sine-type bias
under isotropic tension. Each structure was generated for a bias
parameter \(\alpha = 10\), a Poisson ratio
\(\nu = 0.2\), material strength
\(\sigma_{ct} = \pm 20\) and a fatigue parameter
\(M=10\). The structure for the cosine-bias contains
\(2,000\) particles whereas the structure for the sine-bias
contains \(6,000\) particles.

\par The cosine-type bias leads to very regular needle like cracks (fig. 23)
while the sine-type bias produces a completely
new structure (fig. 24). \par

\par Here, we observe very typical properties. The generated crack is
very similar to the ones observed in the experiments by Lemaire and Van
Damme \SGMPcite{lemaire_vandamme}{}. With a sine-type
bias one obtains a very pronounced ${90}^\circ$ behavior: Initially all side
branches grow perpendicular to their main branches. Like the cracks
produced by Van Damme et al. the branches bend and side branches appear
only on the outer side of the curved crack. Furthermore we observe, that
again the crack tips are very stable and do not show a tip-splitting
instability.\par

\def\XRefId{}\subsection{\SGMPlab\XRefId Quantitative analysis}

All crack patterns shown above have a fractal structure which shall be
analyzed in terms of the dependence of the radius of gyration
\(R_g\) on the crack size \(N\). Therefore we
generated five clusters for each geometry  uniaxial tension,
shear and isotropic tension  and each parameter set  case
1.: high strength \(\sigma_{ct}=\pm 100\) and strong fatigue
\(M=10,000\) and case 2.: low strength
\(\sigma_{ct}=\pm 30\) and weak fatigue \(M=10\). The
results are shown in the figures 25--27. In all cases we find a fractal
behavior
\def\XRefId{}
\begin{equation}\SGMPlab\XRefId\vcenter{\halign{\strut\hfil#\hfil&#\hfil\cr
$\displaystyle{N\propto R^{D_{f}}_{g}}$\cr
}}\end{equation}

where the fractal dimensions \(D_f\) are summarized in the
following table.
\begin{figure} [htbp]

\SGMPbeginTable{}{}{}{}{176:177:178}{C}{}{}{100}
\Rrule[\SGMPTabRuleHS{-:-:-}\mdef\TabRuleH{}]
\Row[\mdef\TabJustV{}\def\TabCellHt{}]
	\Crule[\mdef\TabRuleV{-}]
	\Cell[\TabColSpan=1\TabVSpan=1\mdef\TabJustV{c}
        \mdef\TabJustH{c}\mdef\CellShade{0}]\Endcell[]
	\Crule[\mdef\TabRuleV{-}]
	\Cell[\TabColSpan=1\TabVSpan=1\mdef\TabJustV{c}
        \mdef\TabJustH{c}\mdef\CellShade{0}]case 1.
\Endcell[]
	\Crule[\mdef\TabRuleV{-}]
	\Cell[\TabColSpan=1\TabVSpan=1\mdef\TabJustV{c}
        \mdef\TabJustH{c}\mdef\CellShade{0}]case 2.
\Endcell[]
	\Crule[\mdef\TabRuleV{-}]
\Endrow[]
\Rrule[\SGMPTabRuleHS{-:-:-}\mdef\TabRuleH{}]
\Row[\mdef\TabJustV{}\def\TabCellHt{}]
	\Crule[\mdef\TabRuleV{-}]
	\Cell[\TabColSpan=1\TabVSpan=1\mdef\TabJustV{c}
        \mdef\TabJustH{c}\mdef\CellShade{0}]uniaxial tension
\Endcell[]
	\Crule[\mdef\TabRuleV{-}]
	\Cell[\TabColSpan=1\TabVSpan=1\mdef\TabJustV{c}
         \mdef\TabJustH{c}\mdef\CellShade{0}]1.89{\ifmmode\pm\else$\pm$\fi}0.02
\Endcell[]
	\Crule[\mdef\TabRuleV{-}]
	\Cell[\TabColSpan=1\TabVSpan=1\mdef\TabJustV{c}\mdef\TabJustH{c}
        \mdef\CellShade{0}]1.00{\ifmmode\pm\else$\pm$\fi}0.01
\Endcell[]
	\Crule[\mdef\TabRuleV{-}]
\Endrow[]
\Rrule[\SGMPTabRuleHS{-:-:-}\mdef\TabRuleH{}]
\Row[\mdef\TabJustV{}\def\TabCellHt{}]
	\Crule[\mdef\TabRuleV{-}]
	\Cell[\TabColSpan=1\TabVSpan=1\mdef\TabJustV{c}
\mdef\TabJustH{c}\mdef\CellShade{0}]shear
\Endcell[]
	\Crule[\mdef\TabRuleV{-}]
	\Cell[\TabColSpan=1\TabVSpan=1\mdef\TabJustV{c}
\mdef\TabJustH{c}\mdef\CellShade{0}]1.51{\ifmmode\pm\else$\pm$\fi}0.01
\Endcell[]
	\Crule[\mdef\TabRuleV{-}]
	\Cell[\TabColSpan=1\TabVSpan=1\mdef\TabJustV{c}
\mdef\TabJustH{c}\mdef\CellShade{0}]0.99{\ifmmode\pm\else$\pm$\fi}0.02
\Endcell[]
	\Crule[\mdef\TabRuleV{-}]
\Endrow[]
\Rrule[\SGMPTabRuleHS{-:-:-}\mdef\TabRuleH{}]
\Row[\mdef\TabJustV{}\def\TabCellHt{}]
	\Crule[\mdef\TabRuleV{-}]
	\Cell[\TabColSpan=1\TabVSpan=1\mdef\TabJustV{c}
\mdef\TabJustH{c}\mdef\CellShade{0}]isotropic tension
\Endcell[]
	\Crule[\mdef\TabRuleV{-}]
	\Cell[\TabColSpan=1\TabVSpan=1\mdef\TabJustV{c}
\mdef\TabJustH{c}\mdef\CellShade{0}]1.56{\ifmmode\pm\else$\pm$\fi}0.02
\Endcell[]
	\Crule[\mdef\TabRuleV{-}]
	\Cell[\TabColSpan=1\TabVSpan=1\mdef\TabJustV{c}
\mdef\TabJustH{c}\mdef\CellShade{0}]1.10{\ifmmode\pm\else$\pm$\fi}0.01
\Endcell[]
	\Crule[\mdef\TabRuleV{-}]
\Endrow[]
\Rrule[\SGMPTabRuleHS{-:-:-}\mdef\TabRuleH{}]
\Endtable[]

\end{figure}

The fatigue-parameter \(M\) successfully suppresses side
branching so that the resulting structures are essentially one
dimensional as can be seen from the fractal dimensions in the second
case: here the fractal dimensions are close to unity. In the case of
strong fatigue {---} case 1. {---} the fractal dimensions for shear
and isotropic tension are considerably lower than the fractal dimension
of DLA, which is the result of the observed stability of the crack tips.
Only the fractal dimension of the uniaxial-tension simulations is very
large and indicates that the continuous side branching might lead to a
compact structure.\par

We also measured the fractal dimensions of the simulations using biased
walks. For each case  sine- and cosine-type bias  we
generated five cracks under isotropic tensile load. The results are
shown in figure 28. Again, one observes a fractal
behavior with dimensions
\def\XRefId{}
\begin{equation}\SGMPlab\XRefId\vcenter{\halign{\strut\hfil#\hfil&#\hfil\cr
$\displaystyle{D_{f}=
\left\{\matrix{\SGMPgobble &1.03{\ifmmode\pm\else$\pm$\fi}0.05&\hbox{\rm
cosine-type bias}
\cr \SGMPgobble &1.23{\ifmmode\pm\else$\pm$\fi}0.04&\hbox{\rm sine-type bias}
\cr }\right.}$\cr
}}\end{equation}

One observes that a cosine-type bias leads to one-dimensional,
needle-like structures, whereas the sine-type bias leads to ramified
fractals.

\def\XRefId{Discussion}\section{\SGMPlab\XRefId Discussion}

\par As has been shown above, our new random walker method for crack
growth successfully reproduces well known experimental results and
results of previous small scale simulations using lattice methods.\par

\par The major advantage of our method is the complete absence of any
lattice whatsoever. All three ingredients, the walk, the breaking
criterion and the growth direction, have been formulated completely in a
continuum language. Therefore all lattice effects {---} like the well
known anisotropy in lattice-DLA \SGMPcite{meakin_ball_ramanlal_sander}{} {---}
are avoided. The
observed anisotropies {---} like the pronounced cross-like structure
under shear load {---} have a physical origin and are no artifacts.
This fact is particularly interesting for the observed right angles
between main and side branches. We have to stress again that this
off-lattice method reproduces this important physical property of real
cracks. In fact, we are not aware of other simulations which reproduce
this fact without using a structured lattice.\par

\par An important feature of crack growth is the structural disorder of
real materials. This disorder is implicitly included in our method by
the use of random walkers, which introduces annealed disorder into the
simulation.\par

\par Another advantage of our method is that by using optimized
algorithms \SGMPcite{ossadnik_physica_a_1991}{} all
calculations are easily done on workstations. Actually, our calculations
are all performed on simple SPARC2 stations. They use about 5MB main
memory for a 20,000 particle cluster. The typical run time for one
cluster is in the order of 10{--}15 hours. By altering the strength
and the fatigue parameter this time can be much shorter but also much
longer. The interested reader can obtain all source codes from the
author.\par

\par One disadvantage of our method is that it is not easily possible to
measure the displacement field
\(
\MthAcnt {u}{\matrix{}}\)
 since the whole algorithm is based entirely on the
stress field
\(
\underline{\underline{\SGMPmathgrk{s}}}\).
Therefore it is not yet possible to measure
stress-strain relations or energy release curves, which are of course
important physical quantities. For the same reason it is also not
possible to simulate experimental situations in which not the load but
rather the displacement is controlled. However, one might argue that in
the present stage one is restricted to infinitely stiff materials with
vanishing displacements.\par

\par The last point to be discussed is the fact that {---} except for
the case of the biased walks {---} a walker undergoes an ordinary,
unbiased random walk before touching the cluster. Thus, the probability
that a walker reaches a certain site is implicitly governed by the
Laplace equation. However, this does not impose a wrong, Laplacian
screening behavior, which has two reasons. The first reason for this is,
that the condition for relaunching a walker is completely determined by
the Lam\'e equation and the boundary condition (\SGMPref{boundary}) and, thus,
by the elasticity problem. Furthermore one
has to notice that both, the Laplace and the Lam\'e equation share
the property, that the important fields, the electric and the stress
field, diverge at tips. This is just an explanation of the fact that DLA
clusters {---} like real cracks {---} preferentially grow at tips.
This property is the reason why we chose random walkers to sample the
hot sites of the cracks. In our simulation we can even enhance this tip
growth by strongly reducing the effect of fatigue and thereby inhibiting
the growth of side branches. Thus, the fact that random walks are used
to determine the {``}hot sites{''} of a crack is not in conflict
with our aim to simulate crack growth.\par

\par As an additional possibility we studied the case of biased walks,
in which the bias is calculated from the applied forces. For this case
we have generated cracks which are very similar to experimental ones
observed by Lemaire and Van Damme \SGMPcite{lemaire_vandamme}{}. Although we do
not yet fully
understand the physical meaning of the bias parameter
\(\alpha\) we believe that further research into this direction
will lead to new and interesting results.\par

\def\XRefId{Summary}\section{\SGMPlab\XRefId Summary}

\par In the present work we have studied the possibilities to implement
a novel way to simulate crack growth by using random walkers. Inspired
by Roux we have discussed stochastic algorithms {---} which are
suggested by his original work {---} to solve the Lam\'e
equation. Although he showed an exact relation between random walkers
and elasticity, a direct implementation of the method shown in his work
is not practical, because it is not possible to suppress the strong
fluctuations and to reach useful averages. However, his work showed that
it is indeed possible to treat problems different from Laplacian ones
with random walkers. Inspired by the closely related DLA algorithm
{---} which has been very successful in solving problems of Laplacian
growth {---} we have then constructed a different and new method to
attack the crack growth in elastic media.\par

\par Here, the random walkers are no longer used to construct a Green{'}s
function, but are rather used to detect the hot sites in a crack, which
are the next ones to grow. The formulation of the walk and the growth
rule is purely made in the continuum to avoid all lattice effects. We
are able to reproduce with this new method older results which have been
obtained using other, conventional methods on smaller systems. We are
also able to simulate experiments that have been performed on
concrete.\par

\par Although we restricted ourselves to the special case of
Lam\'e{'}s equation in the framework of linear elasticity, in
principle our method can be extended to other growth phenomena. It may
be interesting to investigate this aspect.\par

\par I would like to thank H. J. Herrmann and S. Roux for many helpful
discussions and comments.\par

\begin{appendix}\def\XRefId{fou_trafo}

\section{Appendix}
Since the matrix
\(\underline{\underline{M}}\)
 in the exponent of (\SGMPref{qtk}) is symmetric, it
can be diagonalized and thus
\(
\underline{\underline{Q}}^{t}{\left\LeftPost{par}\MthAcnt {k}{\matrix{}}
\right\RightPost{par}}\)
 can be evaluated in closed form. One obtains in two
dimensions
\def\XRefId{}
\begin{equation}\SGMPlab\XRefId\vcenter{\halign{\strut\hfil#\hfil&#\hfil\cr
$\displaystyle{\underline{\underline{Q}}^{t}{\left\LeftPost{par}\MthAcnt {k
}{\matrix{}}\right\RightPost{par}}={{1}\over{%
\MthAcnt {k}{\matrix{}}^{2}}}{\left\LeftPost{par}
\matrix{\SGMPgobble &k^{2}_{1}\SGMPlim{r}\exp \SGMPdolim
_{2}+k^{2}_{2}\SGMPlim{r}\exp \SGMPdolim
_{1}&k_{1}k_{2}{\left\LeftPost{cub}
\SGMPlim{r}\exp \SGMPdolim _{2}-\SGMPlim{r}\exp \SGMPdolim
_{1}\right\RightPost{cub}}\cr
\SGMPgobble &k_{1}k_{2}{\left\LeftPost{cub}\SGMPlim{r}\exp \SGMPdolim
_{2}-\SGMPlim{r}\exp \SGMPdolim _{1}
\right\RightPost{cub}}&k^{2}_{1}
\SGMPlim{r}\exp \SGMPdolim _{1}+k^{2}_{%
2}\SGMPlim{r}\exp \SGMPdolim _{2}\cr
}\right\RightPost{par}}}$\cr
}}\end{equation}

where
\(
\SGMPlim{r}\exp \SGMPdolim _{i}=\exp
{\left\LeftPost{par}-t\SGMPmathgrk{l}_{i}\MthAcnt {k}{\matrix{}}^{%
2}\right\RightPost{par}}\)
 and
\(
\SGMPmathgrk{l}_{1}\)
 and
\(
\SGMPmathgrk{l}_{2}\)
 are the eigenvalues
\def\XRefId{}
\begin{equation}\SGMPlab\XRefId\vcenter{\halign{\strut\hfil#\hfil&#\hfil\cr
$\displaystyle{\SGMPmathgrk{l}_{1}={{2\SGMPmathgrk{a}+d}
\over{2d{\left\LeftPost{par}
d+2\right\RightPost{par}}}}}$\cr
$\displaystyle{\SGMPmathgrk{l}_{2}={{3d+2\SGMPmathgrk{a}{\left\LeftPost{par}1-
d\right\RightPost{par}}}\over{2d{\left\LeftPost{par}d+2\right\RightPost{par}}
}}}$\cr
}}\end{equation}

 of
\(
\underline{\underline{M}}\)
. Now, the Fourier transformation into real space
\def\XRefId{}
\begin{equation}\SGMPlab\XRefId\vcenter{\halign{\strut\hfil#\hfil&#\hfil\cr
$\displaystyle{\underline{\underline{Q}}^{t}{\left\LeftPost{par}\MthAcnt {r
}{\matrix{}}\right\RightPost{par}}=\int
\underline{\underline{Q}}^{t}{\left\LeftPost{par}\MthAcnt {k}{\matrix{}}
\right\RightPost{par}}\exp {\left\LeftPost{par}i\MthAcnt {k}{\matrix{}}
\MthAcnt {r}{\matrix{}}\right\RightPost{par}}{{d^{%
2}k}\over{{\left\LeftPost{par}2\SGMPmathgrk{p}\right\RightPost{par}}^{%
2}}}}$\cr
}}\end{equation}

involves the calculation of integrals of the type
\def\XRefId{}
\begin{equation}\SGMPlab\XRefId\vcenter{\halign{\strut\hfil#\hfil&#\hfil\cr
$\displaystyle{\int {{k_{i}k_{j}}
\over{\MthAcnt {k}{\matrix{}}^{2}}}\exp
{\left\LeftPost{par}-t\SGMPmathgrk{l}\MthAcnt {k}{\matrix{}}^{2}
\right\RightPost{par}}\exp {\left\LeftPost{par}i\MthAcnt {k}{\matrix{}}
\MthAcnt {r}{\matrix{}}\right\RightPost{par}}{{d^{%
2}k}\over{{\left\LeftPost{par}2\SGMPmathgrk{p}\right\RightPost{par}}^{%
2}}}.}$\cr
}}\end{equation}

Using polar coordinates
the evaluation of the angular integration leads to Bessel functions of
the first and second kind,
\(
J_{0}{\left\LeftPost{par}x\right\RightPost{par}}\)
 and
\(
J_{2}{\left\LeftPost{par}x\right\RightPost{par}}\)

\def\XRefId{}
\begin{equation}\SGMPlab\XRefId\vcenter{\halign{\strut\hfil#\hfil&#\hfil\cr
$\displaystyle{\int _{0}^{2\SGMPmathgrk{p}}\SGMPlim{r}\cos \SGMPdolim
^{2}\SGMPmathgrk{4}{\ifmmode\cdot\else\.\fi}\exp
{\left\LeftPost{par}ik{\ifmmode\cdot\else\.\fi}r
\cos {\left\LeftPost{par}\SGMPmathgrk{4}-\SGMPmathgrk{4}_{r}
\right\RightPost{par}}
\right\RightPost{par}}\hskip 0.167em d\SGMPmathgrk{4}=
\SGMPmathgrk{p}{\left\LeftPost{sqb}
J_{0}{\left\LeftPost{par}kr\right\RightPost{par}}-\cos
{\left\LeftPost{par}2\SGMPmathgrk{4}_{r}\right\RightPost{par}}J_{%
2}{\left\LeftPost{par}kr\right\RightPost{par}}\right\RightPost{sqb}}
}$\cr
$\displaystyle{\int _{0}^{2\SGMPmathgrk{p}}\SGMPlim{r}\sin \SGMPdolim
^{2}\SGMPmathgrk{4}{\ifmmode\cdot\else\.\fi}
\exp {\left\LeftPost{par}ik{\ifmmode\cdot\else\.\fi}r
\cos {\left\LeftPost{par}\SGMPmathgrk{4}-
\SGMPmathgrk{4}_{r}\right\RightPost{par}}
\right\RightPost{par}}\hskip 0.167em d\SGMPmathgrk{4}=
\SGMPmathgrk{p}{\left\LeftPost{sqb}
J_{0}{\left\LeftPost{par}kr\right\RightPost{par}}+\cos
{\left\LeftPost{par}2\SGMPmathgrk{4}_{r}\right\RightPost{par}}J_{%
2}{\left\LeftPost{par}kr\right\RightPost{par}}\right\RightPost{sqb}}
}$\cr
$\displaystyle{\int _{0}^{2\SGMPmathgrk{p}}\sin
\SGMPmathgrk{4}\cos \SGMPmathgrk{4}{\ifmmode\cdot\else\.\fi}
\exp {\left\LeftPost{par}ik
{\ifmmode\cdot\else\.\fi}r\cos {\left\LeftPost{par}
\SGMPmathgrk{4}-\SGMPmathgrk{4}_{r}
\right\RightPost{par}}\right\RightPost{par}}\hskip 0.167em d\SGMPmathgrk{4}
=\SGMPmathgrk{p}{\left\LeftPost{sqb}-
\sin {\left\LeftPost{par}2\SGMPmathgrk{4}_{%
r}\right\RightPost{par}}J_{2}{\left\LeftPost{par}kr\right\RightPost{par}}
\right\RightPost{sqb}}}$\cr
}}\end{equation}

The remaining radial integration involves an integral of Bessel
functions and Gaussians and finally one obtains \SGMPcite{gradshteyn_ryzhik}{}
\def\XRefId{}
\begin{equation}\SGMPlab\XRefId\vcenter{\halign{\strut\hfil#\hfil&#\hfil\cr
$\displaystyle{\underline{\underline{Q}}^{t}{\left\LeftPost{par}\MthAcnt {r
}{\matrix{}}\right\RightPost{par}}={{1}\over{%
2\SGMPmathgrk{p}}}{\left\LeftPost{sqb}f_{s}{\left\LeftPost{par}r
,t\right\RightPost{par}}+f_{a}{\left\LeftPost{par}r,t\right\RightPost{par}}
\right\RightPost{sqb}}{\ifmmode\cdot\else\.\fi}\underline{\underline{1}}-{{1}
\over{2\SGMPmathgrk{p}}}f_{a}{\left\LeftPost{par}r,t\right\RightPost{par}}
{\ifmmode\cdot\else\.\fi}2{{\MthAcnt {r}{\matrix{}}
\otimes \MthAcnt {r}{\matrix{}}
}\over{\MthAcnt {r}{\matrix{}}^{2}}}
}$\cr
}}\end{equation}

where
\def\XRefId{}
\begin{equation}\SGMPlab\XRefId\vcenter{\halign{\strut\hfil#\hfil&#\hfil\cr
$\displaystyle{f_{s}{\left\LeftPost{par}r,t\right\RightPost{par}}={{%
\exp {\left\LeftPost{par}-r^{2}/4\SGMPmathgrk{l}_{1}
t\right\RightPost{par}}}\over{4\SGMPmathgrk{l}_{1}t}}
+{{\exp {\left\LeftPost{par}-r^{2}/4\SGMPmathgrk{l}_{%
2}t\right\RightPost{par}}}\over{4\SGMPmathgrk{l}_{2}t}}
}$\cr
$\displaystyle{f_{a}{\left\LeftPost{par}r,t\right\RightPost{par}}=
{\left\LeftPost{par}
1+{{4\SGMPmathgrk{l}_{1}t}\over{r^{2}}}
\right\RightPost{par}}{{\exp {\left\LeftPost{par}-r^{%
2}/4\SGMPmathgrk{l}_{1}t\right\RightPost{par}}}\over{4
\SGMPmathgrk{l}_{1}t}}-{\left\LeftPost{par}1+{{4\SGMPmathgrk{l}_{%
2}t}\over{r^{2}}}\right\RightPost{par}}
{{\exp {\left\LeftPost{par}-r^{2}/4\SGMPmathgrk{l}_{%
2}t\right\RightPost{par}}}\over{4\SGMPmathgrk{l}_{2}t}}
}$\cr
}}\end{equation}

\end{appendix}

\end{document}